\documentclass[pre,aps,twocolumn,superscriptaddress,nofootinbib]{revtex4-2}
\usepackage{graphicx}
\usepackage{amssymb,latexsym,amsthm,amsmath}    
\usepackage{mathtools}
\usepackage{float} 
\usepackage[draft]{changes}
\usepackage{hyperref}
\usepackage{color}
\usepackage{hhline}
\usepackage{colortbl}
\usepackage{ulem} 
\usepackage{xcolor}

\begin{document}

\title{Local predictors of explosive synchronization with ordinal methods}
\author{I. Leyva}
\affiliation{Complex Systems Group \& GISC, Universidad Rey Juan Carlos, 28933 M\'ostoles, Spain}
\affiliation{Center for Biomedical Technology, Universidad Polit\'ecnica de Madrid, Pozuelo de Alarc\'on, 28223 Madrid, Spain}

\author{Juan A. Almendral}
\affiliation{Complex Systems Group \& GISC, Universidad Rey Juan Carlos, 28933 M\'ostoles, Spain}
\affiliation{Center for Biomedical Technology, Universidad Polit\'ecnica de Madrid, Pozuelo de Alarc\'on, 28223 Madrid, Spain}

\author{Christophe Letellier}
\affiliation{Rouen Normandie Université-CORIA, Campus Universitaire du Madrillet, F-76800 Saint-Etienne du Rouvray, France}

\author{I. Sendi\~na-Nadal}
\email[Corresponding author:]{irene.sendina@urjc.es}
\affiliation{Complex Systems Group \& GISC, Universidad Rey Juan Carlos, 28933 M\'ostoles, Spain}
\affiliation{Center for Biomedical Technology, Universidad Polit\'ecnica de Madrid, Pozuelo de Alarc\'on, 28223 Madrid, Spain}

\begin{abstract}
We propose to use the ordinal pattern transition (OPT) entropy measured at sentinel central nodes as a potential predictor of explosive transitions to synchronization in networks of various dynamical systems with increasing complexity. Our results demonstrate that the OPT entropic measure surpasses traditional early warning signals (EWS) measures and could be valuable to the tools available for predicting critical transitions. In particular, we investigate networks of diffusively coupled phase oscillators and chaotic R\"ossler systems. 
As maps, we consider a neural network of Chialvo maps coupled in star and scale-free configurations. Furthermore, we apply this measure to time series data obtained from a network of electronic circuits operating in the chaotic regime.
\end{abstract}

\maketitle

\section{Introduction}
Critical transitions, or tipping points, refer to sudden and often irreversible changes in the behavior of systems occurring in various natural and engineered contexts. These phenomena are especially concerning in fields such as ecology \cite{Scheffer2009, Scheffer2012}, neuroscience \cite{Kuhlmann2018, Kim2022, Costa2024}, climate science\cite{Livina2007,Lenton2008,Lenton2011}, and even financial markets, where predicting such transitions is crucial for mitigating adverse outcomes \cite{Moore2018, George2023}. Researchers have extensively studied early warning signals (EWS) to identify impending transitions. These signals include critical slowing down \cite{Patterson2021, Tirabassi2022b, Sardanyes2024, Evers2024}, intermittency \cite{Kalitzin2019, Vera2020b}, and pattern formation \cite{Sardanyes2024, Tirabassi2024}, which appear in many systems.

Complex networks, which describe interconnected systems such as ecosystems, brain networks, and social interactions, are particularly challenging for transition prediction. In these systems, critical transitions often manifest as changes in the collective synchronization state. The shift to a coherent state can enhance the functionality of networked systems, as seen in brain networks \cite{Papo2024} and power grids \cite{Motter2013}. However, this transition to synchronization can also lead to catastrophic consequences, such as epileptic seizures \cite{Kuhlmann2018, Costa2024}, fibromyalgia \cite{Kim2022}, or species extinctions \cite{Aparicio2021, Fan2021}. Understanding how network topology influences system dynamics is essential for predicting synchronization transitions, and several studies have shown that features of the structural \cite{Masuda2024, MacLaren2023} or functional \cite{Ehstand2023} network can serve as indicators for the transition. Identification of sentinel nodes, displaying EWS more prominently than others, has significantly improved prediction accuracy \cite{Aparicio2021, MacLaren2024a}. Sentinel nodes have been utilized in ecological networks to detect potential regime shifts and in brain networks to pinpoint critical regions for neuromodulation \cite{Kim2022, Masuda2024}.

However, traditional EWS measures such as autocorrelation, fluctuation variability, or kurtosis are not universally reliable. In highly nonlinear or stochastic systems, their effectiveness is significantly diminished. For instance, as systems approach critical transitions, higher-order nonlinear effects, often overlooked in linear approximations, become increasingly important  \cite{Bury2021, Kong2021, Vishnoi2024}. Additionally, traditional methods may struggle to distinguish meaningful signals from noise in systems with high-dimensional dynamics, emphasizing the need for more robust techniques \cite{MacLaren2023,MacLaren2024a, Tarigo2024}.

Recently, machine learning (ML) methods have demonstrated the potential to address challenges by identifying complex patterns in data that conventional early warning signals (EWS) may overlook. Studies have demonstrated the utility of ML in predicting tipping points in oscillator ensembles \cite{Bassi2022, Ma2023, George2023, Zhang2022}, ecology systems \cite{Fan2021, Liu2024} or brain networks \cite{Roy2022,  Bassi2022}. Additionally, ordinal methods, which analyze the relative order of data, have emerged as a promising tool for predicting transitions \cite{Shahriari2023, Tirabassi2024, Leyva2020}. By focusing on the sequence of events rather than absolute values, ordinal analysis proves to be robust against noise and non-stationarity, making it well-suited for detecting subtle precursors to changes in evolving systems \cite{Leyva2022, Almendral2023, Tirabassi2023, Lehnertz2023}.

A further challenge in prediction is the occurrence of explosive transitions. This term encompasses various phenomena, such as explosive synchronization (ES) \cite{Boccaletti2016}, sudden structural changes in adaptive systems \cite{Avalos2018}, and abrupt shifts in social dynamics \cite{Soriano2019, Khalil2023}. Explosive transitions have profound implications for real-world systems, as they have been linked to conditions such as epilepsy \cite{Ranjan2024}, hypersensitivity in brain networks \cite{Kim2022}, and massive extinctions \cite{Liu2024}. Predicting these sudden transitions is particularly challenging because collective observables tend to remain stable until the moment of transition. Traditional early warning signals (EWS) often struggle to predict explosive synchronization due to its rapid onset \cite{Karimi2022}. While machine learning approaches have proven helpful in detecting explosive synchronization in small ensembles \cite{Fan2021, Liu2024}, there is still a need for alternative methods that are computationally less demanding.

Recently, the ordinal analysis applied to time series from single nodes of dynamical networks \cite{Tlaie2019, Almendral2023} has revealed a strong correlation between ordinal pattern transition (OPT) entropy and structural and dynamical centrality in a wide range of weakly coupled heterogeneous ensembles. This suggests that ordinal local measures, particularly OPT, can effectively rank the sensitivity of nodes based on their proximity to transitions \cite{MacLaren2023, MacLaren2024a}. This work explores the ordinal transition entropy measured at sentinel central nodes as a potential predictor of explosive transitions to synchronization. Our findings indicate that the OPT entropy adds further information to traditional EWS measures and could be a valuable addition to the tools available for predicting critical transitions.

\section{Methods}

\subsection{Ordinal Patterns and Permutation Entropy Measures}
\label{methods}

Ordinal patterns-based methods are a family of simple and robust techniques that characterize any time series simply by comparing neighbouring relative values \cite{Bandt2002, McCullough2015,Zanin21, Leyva2022}. Given any time series $x=\{x_t; t=1,\dots,T\}$, data are projected into a sequence of symbols of ordinal patterns of a given length $D>1$ obtained from the comparison of consecutive ($\tau=1$) or non-consecutive ($\tau>1$) points in the following manner. First, the time series is divided into blocks $v_t=(x_t,x_{t+\tau},\dots,x_{t+(D-1)\tau})$ of size $D$, and then each element of the block is replaced by a number in $[1,\dots,D]$ corresponding to its relative position when arranged in ascending order.  
In this way, each block $v_t$ is assigned to one of the $D!$ possible orderings 
(permutations) $\pi_{\ell}$ in which the $D$ elements can be ordered. Finally, 
the probability at which the ordinal pattern $\pi_{\ell}$ is found in the time 
series is computed as $p(\pi_{\ell})=\# \{S_t \text{ is mapped into symbol } 
\pi_{\ell}\}/L$, being $L=\lfloor T/(\tau D) \rfloor$ the total number of 
blocks $S_t$ in which the time series is divided ($\lfloor \rfloor$ is the 
floor function). This procedure enables a reliable symbolic dynamics 
mapping a time series to a symbolic sequence of ordinal patterns (where each
sequence of $D$ integers receives a specific symbol). The time series has to be 
sufficiently long, $L\gg D!$, to obtain reliable statistics and a valuable 
ordinal pattern probability distribution $P=\{p(\pi_{\ell}),\ell=1,\dots,D!\}$ \cite{Bandt2002}. 
Throughout this work, we will use $D=3$ (Bandt and Pompe suggest using 
$3\le D\le 7$ for practical purposes \cite{Bandt2002}) and $T=2000$. Although 
the time lag $\tau$ at which the data is sampled is not critical, this 
parameter could be relevant when the system under study has intrinsic time 
scales \cite{zunino2010,soriano2011}. 
Typically, to overcome the dependency of the ordinal pattern probability 
distribution on the sampling time $\tau$, its selection is usually based on 
time delay embedding criteria  \cite{Micco2012,McCullough2015}. In addition to 
using a fix sampling time applied to the original time series, we will also 
consider a Poincar\'e section approach by retaining the relative maxima of the signal \cite{Chrisment2016,Letellier2006}. 

Our goal is to determine whether the probability distribution of ordinal 
patterns associated with the time series of a dynamical network is able to 
detect hidden clues in its collective state, pointing to the occurrence of 
an eventual abrupt phase transition. To achieve this, we will compare two 
entropic quantities estimated from the sequence of the permutation patterns 
defined above: the permutation entropy and the transition permutation entropy.

\subsubsection{Ordinal permutation entropy}

Given the $D$-order permutation probability distribution 
$P=(p(\pi_1),\dots,p(\pi_{D!}))$, the  Bandt-Pompe's permutation entropy is the corresponding Shannon entropy evaluated as
\begin{equation}\label{eqS}
    {\cal S}[P] =-\sum_{\ell=1}^{D!} p_{\ell}\ln{p_{\ell}},
\end{equation}
with the criterion $0^0=1$. We define a normalized permutation entropy as
\begin{equation}\label{eqH}
    {\cal H}[P] = \frac{\cal S}{\cal S_{\rm max}}
\end{equation}
where $\cal S_{\rm max}$ is the Shanon entropy of the uniform permutation probability distribution, that is, ${\cal S}_{\rm max}={\cal S} [P_{\rm e}]$ with $P_{\rm e}=(\frac{1}{D!},\dots,\frac{1}{D!})$. 

\subsubsection{Ordinal permutation transition entropy}

While the previous entropy measure is solely based on the appearance of 
permutation patterns, transitions between patterns may reveal information about 
the finer temporal organization of a dynamical system \cite{Small2013}. We 
define the ordinal transition probability matrix $O_T:=(p_{\ell m} )$ as
\begin{equation}\label{transmat}
    p_{\ell m}=\frac{\#(\pi_{\ell},\pi_m)}{\#(\pi_{\ell})}
\end{equation}
being $p_{\ell m}$ the probability that pattern $m$ follows pattern $\ell$. In 
case $\#(\pi_{\ell})=0$ for some  $\pi_{\ell}$, we assume $p_{\ell m}=0$. We move from having $D!$ patterns to $D!^2$ transitions.
Thus, the time series must be longer to make the transition matrix $O_T$ statistically significant.


Since $\sum_m p_{\ell m}=1$, $O_T$ is a column-stochastic matrix (weighted and directed) whose coefficients are the transition probabilities among ordinal patterns, including self-transitions. 
Thus, we can characterize the dynamics of the transition patterns at the local and global level of a time series. At the local level, we can define the entropy of each pattern $\pi_{\ell}$ through the distribution probability of the patterns that follow it as
\begin{equation}\label{eqHpi}
    {\cal H}_{\pi_{\ell}}=-\frac{1}{ln D!}\sum_{m_=1}^{D!} p_{\ell m}\ln p_{\ell m}.
\end{equation}
which quantifies the 
predictability of the local transitions from the ordinal pattern $\pi_{\ell}$ 
to any other pattern \cite{Masoller2015,McCullough2017}. 

At the global level, we measure the transitional complexity of the whole $O_T$ 
as the average of the local (pattern) entropies ${\cal H}_{\pi{_{\ell}}}$: 
\begin{equation}\label{eqHT}
   {\cal H}_{\rm T} = 
   \frac{1}{D!}\sum_{\ell =1}^{D!}   {\cal H}_{\pi_{\ell}}   
\end{equation}

\subsection{Dynamical networks and local early warning indicators}
\label{sec-measures}

We consider networks whose nodes are either continuous (a flow) or discrete (a 
map) dynamical systems on the phase space. In particular, as an example of 
flows, we will investigate networks of $N$ diffusively coupled phase 
oscillators and chaotic R\"ossler systems, while for maps, we consider a neural 
network of Chialvo maps. The general form of the equations governing the 
dynamics in each case are, respectively, 
\[ \dot{\mathbf{x}}_i=\mathbf{F}_i\left(\mathbf{x}_i\right)
+d \sum_{j=1}^N a_{i j} \mathbf{H}\left(\mathbf{x}_j-\mathbf {x}_i\right) \, ,
\]
where $\mathbf{x}_i(t)$ is the vector state of node $i$ at time $t$,
and $\mathbf{F}_i$ is the local vector field, and for the map,
\[ {\mathbf{x}}_i(t+1)=\mathbf{M}_i\left(\mathbf{x}_i(t)\right)
+d \sum_{j=1}^N a_{ij} \mathbf{H}\left(\mathbf{x}_j(t)-\mathbf{x}_i(t)\right)
\, , \]
where $\mathbf{x}_i(t)$ is a $m$-dimensional vector describing the state of each node $i$ at the $t$-th iteration of the map $\mathbf{M}_i$.
In both cases, $\mathbf{H}$ is the output function describing the units' interaction. The coupling architecture among them is defined by the adjacency matrix $A$ of the graph whose coefficients ${ A}:=(a_{ij})$ are $a_{ij}=1$, if $i$ and $j$ are connected, and $a_{ij}=0$ otherwise, such that, the degree of each node $k_i=\sum_j a_{ij}$.  The parameter $d=\sigma/k_{\rm max}$ acts as a normalized coupling strength by the maximum degree of the network $k_{\rm max}=\max(k_i)$ to compare different realizations of the network.

Explosive phase synchronization appears when the interaction between the network structure and the local node dynamics hinders the formation of microscopic seeds of synchronization as the coupling increases, a mechanism underlying the emergence of a collective coherent state in smooth, second-order transitions. As a result, the system remains incoherent despite very high coupling strengths, only to abruptly transition to a fully synchronized state
\cite{Boccaletti2016}. Most practical ways to create this effect are based on inducing {\it frequency dissasortativity}, ensuring that each node is dynamically isolated \cite{Leyva2013}. When the network is structurally heterogeneous, this dynamical separation is easily achieved by imposing a degree-frequency correlation \cite{Gardenes2011, Leyva2012}. Therefore, in this work, we study highly heterogeneous topologies (star and scale-free (SF) networks), where the nodes are distributed such that their inner temporal scale is proportional to their degree. Details of the dynamical ranges used in each system are provided in their respective Sections. 

As advanced in Sec. \ref{methods}, we propose to use a local entropic measure as an early warning signal of an abrupt transition in a dynamic network. For each node, 
$i$, we consider one of the accessible scalar variables of the vector state, $x_i(t)$, or, when convenient, a function of that variable, and compute the corresponding permutation transition entropy ${\cal H}_{\rm T}^i$. We expect that nodes with the same degree $k$ will have similar dynamical patterns within the network \cite{Tlaie2019}, allowing us to define the averaged $k-$class permutation 
transition entropy
\begin{equation}\label{eqHTk}
   \langle{\cal H}_{\rm T}\rangle_k =  
   \frac{1}{N_k}\sum_{i|k_i=k}  {\cal H}_{\rm T}^i,  
\end{equation}  
where $N_k$ is the number of nodes having degree $k$, and $\langle \rangle_k$ is an ensemble average.  Similarly, we define a $k$-class average for the permutation entropies of those nodes with the same degree $\langle {\cal H}\rangle_k$.

In addition to characterizing the nodal dynamics by the randomness of the ordinal patterns and their transitions, we evaluate the collective state of the dynamical network for increasing values of the control order parameter $d$ by computing the time-averaged phase order parameter,
\begin{equation}\label{eq:R}
R = \frac{1}{N} \langle | \sum_{j=1}^N {\rm e}^{{{\rm i } \theta_j}}  | \rangle_t    
\end{equation}
where $\theta_j$ is the phase of the $j$th oscillator. The phase-order parameter $R$ accounts for the level of phase synchronization ($0\le R\le 1$), and $\langle \rangle_t$ stands for the time average along a sufficiently large time series. 

Finally, to compare the performance of the proposed local entropic measure given by Eq.~(\ref{eqHTk}) with other common early-warning indicators for critical transitions used in the literature \cite{Scheffer2009}, in some cases, we will monitor the amplitude of the fluctuations of each node time series $x_i(t)$, 
$\sigma_{f,i}=\sqrt{\langle x_{i}(t)^2\rangle -\langle x_i(t)\rangle ^2}$
and the normalized autocorrelation 
\begin{equation}
AC_i(l) = \frac{\sum_{t=1}^{T-l} x_i(t) \cdot x_i(t+l)}{\sqrt{\sum_{t=1}^{T-l} x_i^2(t) \cdot \sum_{t=1}^{T-l} x_i^2(t+l)}}
\end{equation}
with lag $l=1$.

\section{Results}

In the following Sections, we present results demonstrating that the transition entropy measured at the hubs of a network is a highly effective metric anticipating the onset of abrupt synchronization transitions in networks of various dynamical systems with increasing complexity. These systems include phase oscillators, a map emulating neuronal firing dynamics, and chaotic amplitude oscillators. Additionally, we apply this measure to experimental time series data obtained from a network of electronic circuits. 

\subsection{Abrupt synchronization transition in  Kuramoto phase oscillators}

\begin{figure}[t!]
  \begin{center}
    \includegraphics[width=\linewidth]{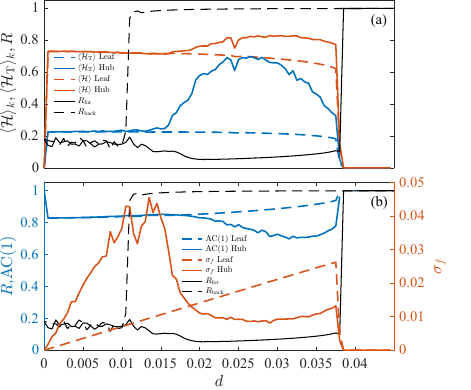}
    \end{center}
  \caption{Abrupt synchronization transition in Kuramoto phase oscillators. (a) $\left<{\cal H}\right>_k$ and $\left<{\cal H}_{\rm T}\right>_k$ as a function of the normalized coupling $d=\sigma/(N-1)$ in a Kuramoto star network of size $N=31$ with $\omega_{o,h}=1.3$ for the hub and $\omega_{o,l}$=1+0.005$\epsilon$ for the leaves with $\epsilon$ a $[0,1]$ random number. Black lines correspond to the Kuramoto order parameter $R$ for the forward (continuous lines) and the backward (dashed lines) transitions. (b) Normalized 1-Lag autocorrelation $AC(1)$ (blue lines, left y-axis for scale) and standard deviation $\sigma_f$ (orange lines, right y-axis for scale) of the signal fluctuations. Input data is a  $\tau=200$ periodic sampling of the instantaneous frequency $\omega_i(t)$ series. Results averaged over 10 random instances.}
\label{fig-kuramoto}
 \end{figure}
 
The first studied system is the Kuramoto network.
\begin{equation}\label{eq-kuram}
    \dot{\theta}_i=\omega_{o,i}+d\sum_{j=1}^N a_{ij} \sin \left(\theta_j-\theta_i\right)
\end{equation}
where the coupling $d$ is the coupling strength and $\omega_{o,i}$ is the natural frequency of node $i$. In the example in Fig. \ref{fig-kuramoto}, we use a 
$N$=31 star-like structure, and to achieve explosive synchronization, we impose a frequency-degree correlation in the nodes \cite{Gardenes2011,Boccaletti2016}, with $\omega_{o,h}=1.3$ for the hub and $\omega_{o,l}=1+0.005\epsilon$ for the leaves with $\epsilon$ a $[0,1]$. 

When searching for signals that indicate an imminent transition, one important decision is determining which observable to monitor, as not all the system outputs are equally sensitive to changes in the state \cite{Letellier2002,Patterson2021,Sendina2022,Masuda2024}. In networked systems, not all nodes are equally effective as predictors of transitions, and their sensitivities may vary due to internal dynamics or their position within the network structure \cite{George2023,MacLaren2024a}. In a Kuramoto network, using the global instantaneous synchronization state $R(t)$ as an early warning signal (EWS) could provide moderate utility \cite{Karimi2022}, but this requires access to a global variable that may not be readily available. In our case, we are more interested in local measures. Therefore, in Fig. \ref{fig-kuramoto}, we use the instantaneous frequency of each oscillator 
$\omega_i(t)$, which is recorded with a sampling time of $\tau = 200$ 
integration steps. 

As seen in ES transitions, the average order parameter $R$ does not provide 
information about the transition and remains close to its initial value until 
it suddenly jumps to its maximum. We then evaluate the ordinal permutation entropy 
$\langle\mathcal{H}\rangle_k$ and the ordinal permutation transition entropy $\langle\mathcal{H}_{\rm T}\rangle_k$ at both 
the $k = N-1$ hub and the $k$ = 1 leaves [Fig. \ref{fig-kuramoto}(a)]. For the 
leaves, the entropy measurements show minor sensitivity to the proximity of the 
transition. However, at the hub, both entropies increase noticeably when the 
coupling is still one-third of its critical synchronization value. Notably, the 
relative increase in $\mathcal{H}_{\rm T}$ is much more significant than that 
of $\mathcal{H}$, providing a clear advantage for generating an early detection 
alarm. Given this advantage, the remaining sections will focus on exploring 
$\mathcal{H}_{\rm T}$ as a more promising measure.

We now compare the result of the local entropy measures with some of those traditionally used as early predictors (Fig. \ref{fig-kuramoto}), the standard deviation of the fluctuations $\sigma_{f,i}$ and the 1-lag autocorrelation AC$_i$(1) of the same data analyzed in panel (a). As for the entropies, none of these EWS are significant for the leaf. However, the dispersion of fluctuations exhibits a linear growth proportional to the coupling that results from the increase in the amplitude of the frequency beat. The AC(1) measure produces a result comparable to $\cal H$ for the hub. On the other hand, the fluctuation dispersion $\sigma_{f}$ of the hub shows a relatively significant maximum (note that the scale of this measure refers to the right Y-axis). However, this maximum does not occur near the synchronous forward transition we are studying, but rather in the region associated with the critical coupling for {\it desynchronization} in the backward transition, a much smaller coupling due to the wide hysteresis. This interesting effect would, however, lead to a false alarm regarding the ES synchronization transition.

\subsection{Abrupt synchronization transitions in coupled Chialvo maps}




\begin{figure*}[t!]
\centering\includegraphics[width=0.75\linewidth]{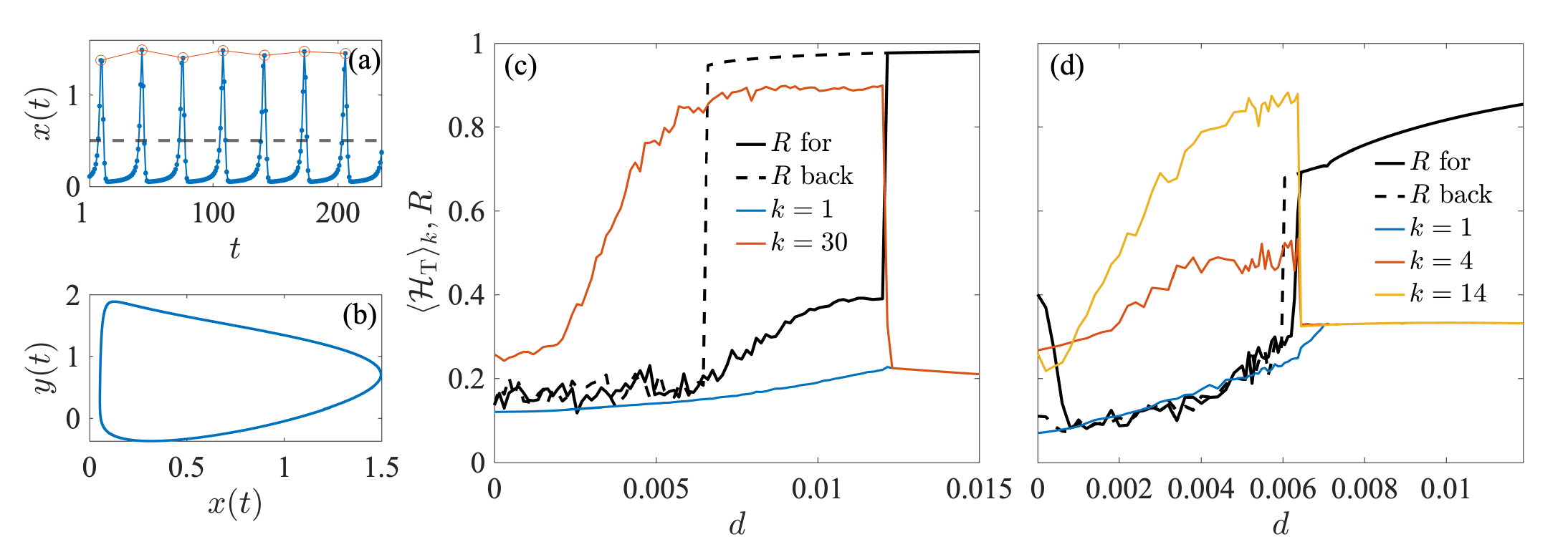}
\caption{Anticipating abrupt synchronization in networks of coupled Chialvo maps. Time series for (a) $x_t$ and phase portrait $x_t-y_t$ of an isolated neural model (\ref{eq-chialvo}) for $a=0.89$, $b=0.6$, $c=0.28$, and $I=0.05$. Red segments connect spike maxima with $x_t>0.5$. (c,d) $\langle {\cal H}_{\rm T}\rangle_k$ for several node degrees (see legend) and Kuramoto order parameter $R$ as a function of the coupling strength $d$ for (c) a star graph of size $N=31$ with $I=0.050$ for the hub and $I=0.049+10^{-4}\epsilon$ for the leaves, with $\epsilon$ a random number uniformly distributed in $[0,1]$, and (d) a scale-free network of $N=100$ nodes of mean degree 4, with $I(k)=0.049+\alpha k$ and $\alpha=3 \cdot 10^{-5}$. In each case, there is an abrupt transition to phase synchronization (black solid line) and an abrupt transition back to desynchronization (black dashed line) with a large hysteresis behavior for the star configuration. 
}
\label{fig-chialvo}
\end{figure*}
To illustrate the performance of the ordinal transition entropy to foresee an upcoming network synchronization event in networks of coupled maps, we consider a neural network of Chialvo maps. The Chialvo map is a two-dimensional neural model that produces periodic or chaotic burst-spike dynamics \cite{chialvo1995}, and it has the following form:
%

\begin{equation}\label{eq-chialvo}
\begin{aligned}
x_i(t+1) &= x^2_i(t) \exp{(y_i(t) - x_i(t))} + I_i + d\sum_{j=1}^N a_{ij}(x_j(t)-x_i(t)),\\[0.01cm]
      y_i(t+1) &= a y_i(t) - b x_i(t) + c 
\end{aligned}
\end{equation}
where $x_i(t)$ acts as a membrane potential and $y_i(t)$ as a restoring variable. The subscripts $t$ represent a discretized time evolution ($t=1,\dots,T$), while subscripts $i$ refer to a node (neuron) of the network ($i=1,\dots,N$). Parameters $a=0.89$, $b=0.6$, and $c=0.28$, are chosen such that the neural dynamics is periodic or quasi-periodic, whose frequency and amplitude is modulated by the parameter $I$, which plays the role of an ion current injected into the neuron. As $I$ increases, the frequency increases monotonically \cite{chialvo1995} (an extensive study of the parameter space of this map is reported in Ref.~\cite{Stankevich2023}). Here, we impose a linear dependence of the constant bias parameter $I$ with the neuron connectivity as $I_i=0.049+\alpha k_i$ 
to induce an explosive synchronization when increasing the control order parameter $d$. 


Figure \ref{fig-chialvo}(a,b) shows an example of the time evolution of the activation variable $x_{t}$ and phase portrait $(x_t,y_t)$ of an isolated neuron for $I_i=0.05$. The red segments connect the spike maxima with $x_t>0.5$, used to define a phase $\theta_{i}(t)$ and be able to monitor the network synchronization as described in Section \ref{sec-measures}. The phase increases by $2\pi$ each time there is a spike, and it is linearly interpolated between spikes \cite{Boccaletti2002} as 
\begin{equation}
\theta_{i}(t)=2 \pi n_i+2 \pi \frac{t-t_{i,n}}{t_{i,n+1}-t_{i,n}}, \quad t_{i,n} \leqslant t<t_{i,n+1}
\end{equation}
where $t_{i,n}$ is the time at which the $i$-th neuron fires its $n_i$-th spike. 

The sequence of maxima is used to construct the ordinal patterns and compute the ordinal transition entropy of Chialvo maps coupled in star and scale-free configurations. Figure \ref{fig-chialvo} shows the evolution of the $k$-class transition entropy $\langle {\cal H}_{\rm T}\rangle_k$ for nodes of increasing degree (colored curves) belonging to a star of size $N=31$ (panel c) and to a scale-free network of size $N=100$ (panel d) as the coupling $d$ increases. In a star configuration, there are only two classes of sensors: the hub and the leaves. In contrast, a scale-free network exhibits a more diverse degree distribution, where many nodes have lower degrees while only a few have significantly larger degrees. The nodes with higher degrees are particularly good candidates for signaling an abrupt phase transition.
As in the Kuramoto case, we observe how, in the two configurations, the largest degree class nodes exhibit a pronounced growth of their transition entropy long before the sudden jump in the order phase parameter $R$ (black continuous line), while the lowest degree ones do not inform about the abrupt change occurring in the network phase synchronization.

\subsection{Predicting explosive transitions in networks of R\"ossler oscillators}
\label{sec-numrossler}

We now consider a dynamical network composed of chaotic oscillators, adding an extra level of complexity to the local dynamics. We chose R\"ossler systems \cite{Ros76} coupled through the $y$ variable, whose governing equations are:

\begin{equation}
\begin{aligned}
\dot{x}_i&=-w_i y_i-z_i, \\
\dot{y}_i&=w_i x_i+a y_i + d\sum_{j=1}^N a_{ij}(y_j-y_i) , \\
\dot{z}_i&=b +  z_i(x_i-c),
\end{aligned}
\end{equation}
where $a=0.165$, $b=0.4$, $c=8.5$, are set to get a phase-coherent chaotic attractor, and $w_i$ is a control parameter that directly tunes the main frequency of the chaotic oscillator. We set the linear frequency-degree correlation as $\omega_i(k)=1.06+\alpha k_i$ with $\alpha=2.73 \cdot 10^{-4}$ to induce an explosive synchronization transition. To monitor the network state of phase synchronization, we define the phase of each R\"ossler oscillator as $\theta_i=\arctan{(y_i/x_i)}$ and compute the phase order parameter as given by Eq.~(\ref{eq:R}).

\begin{figure*}[t!]
    \centering
    \includegraphics[width=0.75\linewidth]{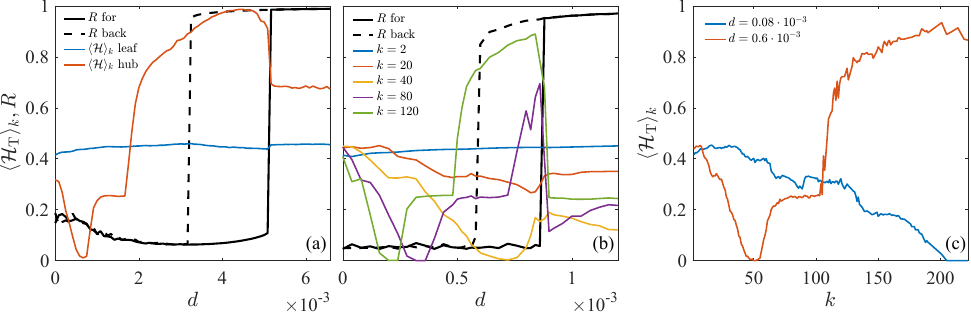}
    \caption{Predicting explosive transitions in networks of R\"ossler oscillators. (a,b) $\langle {\cal H}_{\rm T} \rangle_k$ for different $k$ values (colored lines) and order parameter $R$ for the forward (black continuous line) and backward (black dashed line) routes to explosive synchronization/desynchronization as the coupling strength $d$ increases/decreases for (a) a star of size $N=31$ and (b) a $N=500$ scale-free network ($\langle k\rangle=4$, $\gamma=2.25$). (c) $\langle {\cal H}_{\rm T} \rangle_k$ as a function of $k$ for two values of the coupling strength in (b): a relatively small coupling value, $d=0.08\cdot 10^{-3}$ (blue line), and a value still far from the transition, $d=0.6\cdot 10^{-3}\ll d_{\rm t}\sim 0.9\cdot 10^{-3}$ (red line). In (b) and (c), each point averages the result of 64 independent network realizations.}
    \label{fig:rossler}
\end{figure*}

To investigate this dynamical network, we chose to map each node's vector state $\mathbf{x}_i=(x_i,y_i,z_i)$ to the one-dimensional time series $y_i(t_m), m=1,\dots,T$ generated by the Poincar\'e section $
\mathcal{P} \equiv\left\{\left[x_i\left(t_m\right), z_i\left(t_m\right)\right] \in \mathbb{R}^2 \mid \dot{y}_i\left(t_m\right)=0, \ddot{y}_i\left(t_m\right)>0\right\}$\cite{Shahriari2023}, that is, we take the minima of the $y$ time series to build the $D=3$ permutation patterns.  

Figure \ref{fig:rossler} condenses the main results, evidencing again how the hubs'    transition entropy ${\cal H}_{\rm T}$ 
 detect an explosive synchronization event in advance while the phase order parameter $R$ does not sense any collective effect. In detail, Fig.~\ref{fig:rossler}(a) shows a markedly distinct behavior of the OPT entropies for the leaf and hub of a star network. While the $\langle {\cal H}_{\rm T}\rangle$ of the leaf remains almost flat along the synchronization path, the hub's entropy rises at coupling values even before the hysteresis window starts, defined by the continuous and dashed $R$ curves for increasing ($R_{\rm for}$) and decreasing ($R_{\rm back}$) strength of the coupling. In more complex situations with a heterogeneous degree distribution, as shown in Fig.~\ref{fig:rossler}(b), we observe a similar hierarchical trend where nodes in higher degree classes begin to gain entropy at smaller coupling values. Figure ~\ref{fig:rossler}(c) shows the sensitivity of the OTP entropy as a function of the node's degree depending on the distance from the abrupt transition. While for very low coupling values ($d=0.08\cdot 10^{-3}$), the $\langle {\cal H}_{\rm T}\rangle_k$ monotonously decreases with $k$, for closer values to the hysteresis window ($d=0.6\cdot 10^{-3}$) there is a clear cutoff around $k_{\rm c}\sim 50$ indicating that nodes with $k>k_{\rm c}$ are potential EWS sensors.

\subsection{Predicting explosive transitions in networks of electronic circuits}

Finally, we test the performance of the OPT entropy $\mathcal{H}_{\rm T}$ as an early warning measure for an experimental case of explosive synchronization. The experiment consists of an $N=6$ star network of piecewise R\"ossler electronic circuits operating in the chaotic regime, as the numerical counterpart in the previous Section. The circuits are configured such that the central node oscillates with a mean frequency of 3333 Hz, and the leaf nodes are set with frequencies in the range of 2240 $\pm$ 200 Hz. The transition to synchrony of this ensemble was first analyzed in Ref. \cite{Leyva2012}, where the rest of the experimental details can be found. 

\begin{figure}[t!]
  \begin{center}
    \includegraphics[width=\linewidth]{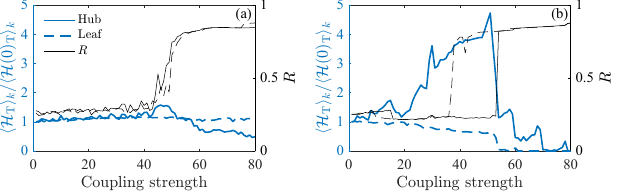}
    \end{center}
  \caption{Normalized ordinal pattern transition entropies $\left<{\cal H}_{\rm T}\right>_k/\left<{\cal H}(0)_{\rm T}\right>_k$ (left y-axis) and phase order parameter $R$ (right y-axis) as a function of the coupling strength in a star network of $N=6$ piecewise R\"ossler electronic circuits for the case of (a) continuous and (b) explosive transitions to synchronization. Input data are Poincar\'e sections series of a voltage local maxima.}
\label{fig-experiment}
\end{figure}

In the numerical analysis presented in Section \ref{sec-numrossler}, the node coupling and the variable being analyzed correspond to $y_i$, while in the experiment, this role is assigned to $x$. To align our analysis with the results above, we extract the Poincar\`e section from the experimental data to generate the $D$=3 permutation patterns. An experimental parameter controls the node's dynamics so that transition can be tuned from second-order (Fig. \ref{fig-experiment}(a))  to explosive (Fig. \ref{fig-experiment}(b))  without affecting the node's frequencies or the chaotic state. The phase reconstruction and synchronization monitoring are performed as in Sec. \ref{sec-numrossler}.

For a better comparison, in Fig. \ref{fig-experiment} we normalize the OPT entropies of each node  $\left<\cal{H_{\rm \rm T}}\right>_k$ to their respective value when the system is uncoupled, $\left<\cal{H({\rm 0})_{\rm T}}\right>_k$. In the continuous transition (a), neither the hub nor the leaves show changes in their initial entropy value along the synchronization process. In contrast, during the ES transition plotted in Fig. \ref{fig-experiment}(b), the hub OPT entropy increases by up to a factor of 5 when the coupling value is still far from the transition. Meanwhile, the leaf experiences a slight reduction before the transition. This result is similar to the one obtained in Fig. \ref{fig:rossler}(a) for the numerical case, which confirms the robustness of this ordinal method as an EWS for the explosive transition even in this case subject to the node's experimental heterogeneity and noisy environment. 

\section{Conclusions}

Predicting state transitions in complex systems is a difficult challenge due to the enormous variety of systems, forms of transitions, and diversity of data. In the case of abrupt transitions, it is even more difficult since many of the system's observables do not show changes during the stages before the transition. In this work, we explore the efficacy of ordinal pattern transition (OPT) entropy as a predictive tool for explosive synchronization transitions in various dynamical networks. Through simulations and experimental validations, OPT entropy can identify subtle signals of the proximity of the transitions across diverse network configurations and dynamical regimes, even outperforming traditional early warning signals (EWS). 

Our results indicate that the sensitivity of OPT entropy is particularly strong in central nodes, which can function as sentinel nodes based on this observable measure. We successfully applied this method to chaotic networks of electronic circuits, demonstrating its versatility and practical utility in real-world applications. These findings highlight the potential of incorporating ordinal methods into the toolkit for studying critical transitions in complex systems, especially when traditional early warning system (EWS) measures struggle under high dimensionality or nonlinearity conditions.

Future research could focus on refining ordinal measures to enhance their sensitivity and applicability to larger and more diverse networks. Additionally, integrating machine learning with ordinal methods may prove beneficial in extending their predictive capabilities across various fields such as neuroscience, ecology, and engineering. This study paves the way for broader applications of ordinal methodologies in forecasting and mitigating the effects of critical transitions in complex dynamical systems.

\section*{Acknowledgments}
This research was supported by the Spanish  Ministerio de Ciencia e Innovación under Projects PID2020-113737GB-I00 and PID2023-147827NB-I00.

%


\begin{thebibliography}{67}%
\makeatletter
\providecommand \@ifxundefined [1]{%
 \@ifx{#1\undefined}
}%
\providecommand \@ifnum [1]{%
 \ifnum #1\expandafter \@firstoftwo
 \else \expandafter \@secondoftwo
 \fi
}%
\providecommand \@ifx [1]{%
 \ifx #1\expandafter \@firstoftwo
 \else \expandafter \@secondoftwo
 \fi
}%
\providecommand \natexlab [1]{#1}%
\providecommand \enquote  [1]{``#1''}%
\providecommand \bibnamefont  [1]{#1}%
\providecommand \bibfnamefont [1]{#1}%
\providecommand \citenamefont [1]{#1}%
\providecommand \href@noop [0]{\@secondoftwo}%
\providecommand \href [0]{\begingroup \@sanitize@url \@href}%
\providecommand \@href[1]{\@@startlink{#1}\@@href}%
\providecommand \@@href[1]{\endgroup#1\@@endlink}%
\providecommand \@sanitize@url [0]{\catcode `\\12\catcode `\$12\catcode
  `\&12\catcode `\#12\catcode `\^12\catcode `\_12\catcode `\%12\relax}%
\providecommand \@@startlink[1]{}%
\providecommand \@@endlink[0]{}%
\providecommand \url  [0]{\begingroup\@sanitize@url \@url }%
\providecommand \@url [1]{\endgroup\@href {#1}{\urlprefix }}%
\providecommand \urlprefix  [0]{URL }%
\providecommand \Eprint [0]{\href }%
\providecommand \doibase [0]{https://doi.org/}%
\providecommand \selectlanguage [0]{\@gobble}%
\providecommand \bibinfo  [0]{\@secondoftwo}%
\providecommand \bibfield  [0]{\@secondoftwo}%
\providecommand \translation [1]{[#1]}%
\providecommand \BibitemOpen [0]{}%
\providecommand \bibitemStop [0]{}%
\providecommand \bibitemNoStop [0]{.\EOS\space}%
\providecommand \EOS [0]{\spacefactor3000\relax}%
\providecommand \BibitemShut  [1]{\csname bibitem#1\endcsname}%
\let\auto@bib@innerbib\@empty
\bibitem [{\citenamefont {Scheffer}\ \emph {et~al.}(2009)\citenamefont
  {Scheffer}, \citenamefont {Bascompte}, \citenamefont {Brock}, \citenamefont
  {Brovkin}, \citenamefont {Carpenter}, \citenamefont {Dakos}, \citenamefont
  {Held}, \citenamefont {Nes}, \citenamefont {Rietkerk},\ and\ \citenamefont
  {Sugihara}}]{Scheffer2009}%
  \BibitemOpen
  \bibfield  {author} {\bibinfo {author} {\bibfnamefont {M.}~\bibnamefont
  {Scheffer}}, \bibinfo {author} {\bibfnamefont {J.}~\bibnamefont {Bascompte}},
  \bibinfo {author} {\bibfnamefont {W.~A.}\ \bibnamefont {Brock}}, \bibinfo
  {author} {\bibfnamefont {V.}~\bibnamefont {Brovkin}}, \bibinfo {author}
  {\bibfnamefont {S.~R.}\ \bibnamefont {Carpenter}}, \bibinfo {author}
  {\bibfnamefont {V.}~\bibnamefont {Dakos}}, \bibinfo {author} {\bibfnamefont
  {H.}~\bibnamefont {Held}}, \bibinfo {author} {\bibfnamefont {E.~H.~V.}\
  \bibnamefont {Nes}}, \bibinfo {author} {\bibfnamefont {M.}~\bibnamefont
  {Rietkerk}},\ and\ \bibinfo {author} {\bibfnamefont {G.}~\bibnamefont
  {Sugihara}},\ }\bibfield  {title} {\bibinfo {title} {Early-warning signals
  for critical transitions},\ }\href {https://doi.org/10.1038/nature08227}
  {\bibfield  {journal} {\bibinfo  {journal} {Nature}\ }\textbf {\bibinfo
  {volume} {461}},\ \bibinfo {pages} {53} (\bibinfo {year} {2009})}\BibitemShut
  {NoStop}%
\bibitem [{\citenamefont {Scheffer}\ \emph {et~al.}(2012)\citenamefont
  {Scheffer}, \citenamefont {Carpenter}, \citenamefont {Lenton}, \citenamefont
  {Bascompte}, \citenamefont {Brock}, \citenamefont {Dakos}, \citenamefont
  {van~de Koppel}, \citenamefont {van~de Leemput}, \citenamefont {Levin},
  \citenamefont {van Nes}, \citenamefont {Pascual},\ and\ \citenamefont
  {Vandermeer}}]{Scheffer2012}%
  \BibitemOpen
  \bibfield  {author} {\bibinfo {author} {\bibfnamefont {M.}~\bibnamefont
  {Scheffer}}, \bibinfo {author} {\bibfnamefont {S.~R.}\ \bibnamefont
  {Carpenter}}, \bibinfo {author} {\bibfnamefont {T.~M.}\ \bibnamefont
  {Lenton}}, \bibinfo {author} {\bibfnamefont {J.}~\bibnamefont {Bascompte}},
  \bibinfo {author} {\bibfnamefont {W.}~\bibnamefont {Brock}}, \bibinfo
  {author} {\bibfnamefont {V.}~\bibnamefont {Dakos}}, \bibinfo {author}
  {\bibfnamefont {J.}~\bibnamefont {van~de Koppel}}, \bibinfo {author}
  {\bibfnamefont {I.~A.}\ \bibnamefont {van~de Leemput}}, \bibinfo {author}
  {\bibfnamefont {S.~A.}\ \bibnamefont {Levin}}, \bibinfo {author}
  {\bibfnamefont {E.~H.}\ \bibnamefont {van Nes}}, \bibinfo {author}
  {\bibfnamefont {M.}~\bibnamefont {Pascual}},\ and\ \bibinfo {author}
  {\bibfnamefont {J.}~\bibnamefont {Vandermeer}},\ }\bibfield  {title}
  {\bibinfo {title} {Anticipating critical transitions},\ }\href
  {https://doi.org/10.1126/science.1225244} {\bibfield  {journal} {\bibinfo
  {journal} {Science}\ }\textbf {\bibinfo {volume} {338}},\ \bibinfo {pages}
  {344} (\bibinfo {year} {2012})}\BibitemShut {NoStop}%
\bibitem [{\citenamefont {Kuhlmann}\ \emph {et~al.}(2018)\citenamefont
  {Kuhlmann}, \citenamefont {Lehnertz}, \citenamefont {Richardson},
  \citenamefont {Schelter},\ and\ \citenamefont {Zaveri}}]{Kuhlmann2018}%
  \BibitemOpen
  \bibfield  {author} {\bibinfo {author} {\bibfnamefont {L.}~\bibnamefont
  {Kuhlmann}}, \bibinfo {author} {\bibfnamefont {K.}~\bibnamefont {Lehnertz}},
  \bibinfo {author} {\bibfnamefont {M.~P.}\ \bibnamefont {Richardson}},
  \bibinfo {author} {\bibfnamefont {B.}~\bibnamefont {Schelter}},\ and\
  \bibinfo {author} {\bibfnamefont {H.~P.}\ \bibnamefont {Zaveri}},\ }\bibfield
   {title} {\bibinfo {title} {Seizure prediction — ready for a new era},\
  }\href {https://doi.org/10.1038/s41582-018-0055-2} {\bibfield  {journal}
  {\bibinfo  {journal} {Nature Reviews Neurology}\ }\textbf {\bibinfo {volume}
  {14}},\ \bibinfo {pages} {618} (\bibinfo {year} {2018})}\BibitemShut
  {NoStop}%
\bibitem [{\citenamefont {Kim}\ \emph {et~al.}(2022)\citenamefont {Kim},
  \citenamefont {Harris}, \citenamefont {DaSilva},\ and\ \citenamefont
  {Lee}}]{Kim2022}%
  \BibitemOpen
  \bibfield  {author} {\bibinfo {author} {\bibfnamefont {M.~K.}\ \bibnamefont
  {Kim}}, \bibinfo {author} {\bibfnamefont {R.~E.}\ \bibnamefont {Harris}},
  \bibinfo {author} {\bibfnamefont {A.~F.}\ \bibnamefont {DaSilva}},\ and\
  \bibinfo {author} {\bibfnamefont {U.~C.}\ \bibnamefont {Lee}},\ }\bibfield
  {title} {\bibinfo {title} {Explosive synchronization-based brain modulation
  reduces hypersensitivity in the brain network: A computational model study},\
  }\bibfield  {journal} {\bibinfo  {journal} {Frontiers in Computational
  Neuroscience}\ }\textbf {\bibinfo {volume} {16}},\ \href
  {https://doi.org/10.3389/fncom.2022.815099} {10.3389/fncom.2022.815099}
  (\bibinfo {year} {2022})\BibitemShut {NoStop}%
\bibitem [{\citenamefont {Costa}\ \emph {et~al.}(2024)\citenamefont {Costa},
  \citenamefont {Teixeira},\ and\ \citenamefont {Pinto}}]{Costa2024}%
  \BibitemOpen
  \bibfield  {author} {\bibinfo {author} {\bibfnamefont {G.}~\bibnamefont
  {Costa}}, \bibinfo {author} {\bibfnamefont {C.}~\bibnamefont {Teixeira}},\
  and\ \bibinfo {author} {\bibfnamefont {M.~F.}\ \bibnamefont {Pinto}},\
  }\bibfield  {title} {\bibinfo {title} {Comparison between epileptic seizure
  prediction and forecasting based on machine learning},\ }\href
  {https://doi.org/10.1038/s41598-024-56019-z} {\bibfield  {journal} {\bibinfo
  {journal} {Scientific Reports}\ }\textbf {\bibinfo {volume} {14}},\ \bibinfo
  {pages} {5653} (\bibinfo {year} {2024})}\BibitemShut {NoStop}%
\bibitem [{\citenamefont {Livina}\ and\ \citenamefont
  {Lenton}(2007)}]{Livina2007}%
  \BibitemOpen
  \bibfield  {author} {\bibinfo {author} {\bibfnamefont {V.~N.}\ \bibnamefont
  {Livina}}\ and\ \bibinfo {author} {\bibfnamefont {T.~M.}\ \bibnamefont
  {Lenton}},\ }\bibfield  {title} {\bibinfo {title} {A modified method for
  detecting incipient bifurcations in a dynamical system},\ }\href
  {https://doi.org/10.1029/2006GL028672} {\bibfield  {journal} {\bibinfo
  {journal} {Geophysical Research Letters}\ }\textbf {\bibinfo {volume} {34}},\
  \bibinfo {pages} {L03712} (\bibinfo {year} {2007})}\BibitemShut {NoStop}%
\bibitem [{\citenamefont {Lenton}\ \emph {et~al.}(2008)\citenamefont {Lenton},
  \citenamefont {Held}, \citenamefont {Kriegler}, \citenamefont {Hall},
  \citenamefont {Lucht}, \citenamefont {Rahmstorf},\ and\ \citenamefont
  {Schellnhuber}}]{Lenton2008}%
  \BibitemOpen
  \bibfield  {author} {\bibinfo {author} {\bibfnamefont {T.~M.}\ \bibnamefont
  {Lenton}}, \bibinfo {author} {\bibfnamefont {H.}~\bibnamefont {Held}},
  \bibinfo {author} {\bibfnamefont {E.}~\bibnamefont {Kriegler}}, \bibinfo
  {author} {\bibfnamefont {J.~W.}\ \bibnamefont {Hall}}, \bibinfo {author}
  {\bibfnamefont {W.}~\bibnamefont {Lucht}}, \bibinfo {author} {\bibfnamefont
  {S.}~\bibnamefont {Rahmstorf}},\ and\ \bibinfo {author} {\bibfnamefont
  {H.~J.}\ \bibnamefont {Schellnhuber}},\ }\bibfield  {title} {\bibinfo {title}
  {Tipping elements in the {Ea}rth's climate system},\ }\href
  {https://doi.org/10.1073/pnas.0705414105} {\bibfield  {journal} {\bibinfo
  {journal} {Proceedings of the National Academy of Sciences}\ }\textbf
  {\bibinfo {volume} {105}},\ \bibinfo {pages} {1786} (\bibinfo {year}
  {2008})}\BibitemShut {NoStop}%
\bibitem [{\citenamefont {Lenton}(2011)}]{Lenton2011}%
  \BibitemOpen
  \bibfield  {author} {\bibinfo {author} {\bibfnamefont {T.}~\bibnamefont
  {Lenton}},\ }\bibfield  {title} {\bibinfo {title} {Early warning of climate
  tipping points},\ }\href {https://doi.org/10.1038/nclimate1143} {\bibfield
  {journal} {\bibinfo  {journal} {Nature Climate Change}\ }\textbf {\bibinfo
  {volume} {1}},\ \bibinfo {pages} {201} (\bibinfo {year} {2011})}\BibitemShut
  {NoStop}%
\bibitem [{\citenamefont {Moore}(2018)}]{Moore2018}%
  \BibitemOpen
  \bibfield  {author} {\bibinfo {author} {\bibfnamefont {J.~C.}\ \bibnamefont
  {Moore}},\ }\bibfield  {title} {\bibinfo {title} {Predicting tipping points
  in complex environmental systems},\ }\href
  {https://doi.org/10.1073/pnas.1721206115} {\bibfield  {journal} {\bibinfo
  {journal} {Proceedings of the National Academy of Sciences}\ }\textbf
  {\bibinfo {volume} {115}},\ \bibinfo {pages} {635} (\bibinfo {year}
  {2018})}\BibitemShut {NoStop}%
\bibitem [{\citenamefont {George}\ \emph {et~al.}(2023)\citenamefont {George},
  \citenamefont {Kachhara},\ and\ \citenamefont {Ambika}}]{George2023}%
  \BibitemOpen
  \bibfield  {author} {\bibinfo {author} {\bibfnamefont {S.~V.}\ \bibnamefont
  {George}}, \bibinfo {author} {\bibfnamefont {S.}~\bibnamefont {Kachhara}},\
  and\ \bibinfo {author} {\bibfnamefont {G.}~\bibnamefont {Ambika}},\
  }\bibfield  {title} {\bibinfo {title} {Early warning signals for critical
  transitions in complex systems},\ }\href
  {https://doi.org/10.1088/1402-4896/acde20} {\bibfield  {journal} {\bibinfo
  {journal} {Physica Scripta}\ }\textbf {\bibinfo {volume} {98}},\ \bibinfo
  {pages} {072002} (\bibinfo {year} {2023})}\BibitemShut {NoStop}%
\bibitem [{\citenamefont {Patterson}\ \emph {et~al.}(2021)\citenamefont
  {Patterson}, \citenamefont {Strang},\ and\ \citenamefont
  {Abbott}}]{Patterson2021}%
  \BibitemOpen
  \bibfield  {author} {\bibinfo {author} {\bibfnamefont {A.~C.}\ \bibnamefont
  {Patterson}}, \bibinfo {author} {\bibfnamefont {A.~G.}\ \bibnamefont
  {Strang}},\ and\ \bibinfo {author} {\bibfnamefont {K.~C.}\ \bibnamefont
  {Abbott}},\ }\bibfield  {title} {\bibinfo {title} {When and where we can
  expect to see early warning signals in multispecies systems approaching
  tipping points: Insights from theory},\ }\href
  {https://doi.org/10.1086/714275} {\bibfield  {journal} {\bibinfo  {journal}
  {American Naturalist}\ }\textbf {\bibinfo {volume} {198}},\ \bibinfo {pages}
  {E12} (\bibinfo {year} {2021})}\BibitemShut {NoStop}%
\bibitem [{\citenamefont {Tirabassi}\ and\ \citenamefont
  {Masoller}(2022)}]{Tirabassi2022b}%
  \BibitemOpen
  \bibfield  {author} {\bibinfo {author} {\bibfnamefont {G.}~\bibnamefont
  {Tirabassi}}\ and\ \bibinfo {author} {\bibfnamefont {C.}~\bibnamefont
  {Masoller}},\ }\bibfield  {title} {\bibinfo {title} {Correlation lags give
  early warning signals of approaching bifurcations},\ }\bibfield  {journal}
  {\bibinfo  {journal} {Chaos, Solitons \& Fractals}\ }\textbf {\bibinfo
  {volume} {155}},\ \href {https://doi.org/10.1016/j.chaos.2021.111720}
  {10.1016/j.chaos.2021.111720} (\bibinfo {year} {2022})\BibitemShut {NoStop}%
\bibitem [{\citenamefont {Sardanyés}\ \emph {et~al.}(2024)\citenamefont
  {Sardanyés}, \citenamefont {Ivančić},\ and\ \citenamefont
  {Vidiella}}]{Sardanyes2024}%
  \BibitemOpen
  \bibfield  {author} {\bibinfo {author} {\bibfnamefont {J.}~\bibnamefont
  {Sardanyés}}, \bibinfo {author} {\bibfnamefont {F.}~\bibnamefont
  {Ivančić}},\ and\ \bibinfo {author} {\bibfnamefont {B.}~\bibnamefont
  {Vidiella}},\ }\bibfield  {title} {\bibinfo {title} {Identifying regime
  shifts, transients and late warning signals for proactive ecosystem
  management},\ }\href {https://doi.org/10.1016/j.biocon.2023.110433}
  {\bibfield  {journal} {\bibinfo  {journal} {Biological Conservation}\
  }\textbf {\bibinfo {volume} {290}},\ \bibinfo {pages} {110433} (\bibinfo
  {year} {2024})}\BibitemShut {NoStop}%
\bibitem [{\citenamefont {Evers}\ \emph {et~al.}(2024)\citenamefont {Evers},
  \citenamefont {Borsboom}, \citenamefont {Fried}, \citenamefont {Hasselman},\
  and\ \citenamefont {Waldorp}}]{Evers2024}%
  \BibitemOpen
  \bibfield  {author} {\bibinfo {author} {\bibfnamefont {K.}~\bibnamefont
  {Evers}}, \bibinfo {author} {\bibfnamefont {D.}~\bibnamefont {Borsboom}},
  \bibinfo {author} {\bibfnamefont {E.~I.}\ \bibnamefont {Fried}}, \bibinfo
  {author} {\bibfnamefont {F.}~\bibnamefont {Hasselman}},\ and\ \bibinfo
  {author} {\bibfnamefont {L.}~\bibnamefont {Waldorp}},\ }\bibfield  {title}
  {\bibinfo {title} {Early warning signals of complex critical transitions in
  deterministic dynamics},\ }\href {https://doi.org/10.1007/s11071-024-10023-0}
  {\bibfield  {journal} {\bibinfo  {journal} {Nonlinear Dynamics}\ }\textbf
  {\bibinfo {volume} {112}},\ \bibinfo {pages} {19071} (\bibinfo {year}
  {2024})}\BibitemShut {NoStop}%
\bibitem [{\citenamefont {Kalitzin}\ \emph {et~al.}(2019)\citenamefont
  {Kalitzin}, \citenamefont {Petkov}, \citenamefont {Suffczynski},
  \citenamefont {Grigorovsky}, \citenamefont {Bardakjian}, \citenamefont
  {Silva},\ and\ \citenamefont {Carlen}}]{Kalitzin2019}%
  \BibitemOpen
  \bibfield  {author} {\bibinfo {author} {\bibfnamefont {S.}~\bibnamefont
  {Kalitzin}}, \bibinfo {author} {\bibfnamefont {G.}~\bibnamefont {Petkov}},
  \bibinfo {author} {\bibfnamefont {P.}~\bibnamefont {Suffczynski}}, \bibinfo
  {author} {\bibfnamefont {V.}~\bibnamefont {Grigorovsky}}, \bibinfo {author}
  {\bibfnamefont {B.~L.}\ \bibnamefont {Bardakjian}}, \bibinfo {author}
  {\bibfnamefont {F.~L.~D.}\ \bibnamefont {Silva}},\ and\ \bibinfo {author}
  {\bibfnamefont {P.~L.}\ \bibnamefont {Carlen}},\ }\bibfield  {title}
  {\bibinfo {title} {Epilepsy as a manifestation of a multistate network of
  oscillatory systems},\ }\href {https://doi.org/10.1016/j.nbd.2019.104488}
  {\bibfield  {journal} {\bibinfo  {journal} {Neurobiology of Disease}\ ,\
  \bibinfo {pages} {104488}} (\bibinfo {year} {2019})}\BibitemShut {NoStop}%
\bibitem [{\citenamefont {Vera-Ávila}\ \emph {et~al.}(2020)\citenamefont
  {Vera-Ávila}, \citenamefont {Sevilla-Escoboza},\ and\ \citenamefont
  {Leyva}}]{Vera2020b}%
  \BibitemOpen
  \bibfield  {author} {\bibinfo {author} {\bibfnamefont {V.~P.}\ \bibnamefont
  {Vera-Ávila}}, \bibinfo {author} {\bibfnamefont {J.~R.}\ \bibnamefont
  {Sevilla-Escoboza}},\ and\ \bibinfo {author} {\bibfnamefont {I.}~\bibnamefont
  {Leyva}},\ }\bibfield  {title} {\bibinfo {title} {Complex networks exhibit
  intermittent synchronization},\ }\href {https://doi.org/10.1063/5.0020419}
  {\bibfield  {journal} {\bibinfo  {journal} {Chaos}\ }\textbf {\bibinfo
  {volume} {30}},\ \bibinfo {pages} {103119} (\bibinfo {year}
  {2020})}\BibitemShut {NoStop}%
\bibitem [{\citenamefont {Tirabassi}(2024)}]{Tirabassi2024}%
  \BibitemOpen
  \bibfield  {author} {\bibinfo {author} {\bibfnamefont {G.}~\bibnamefont
  {Tirabassi}},\ }\bibfield  {title} {\bibinfo {title} {Linear theory of the
  spatial signatures of critical slowing down},\ }\href
  {https://doi.org/10.1103/PhysRevResearch.6.023228} {\bibfield  {journal}
  {\bibinfo  {journal} {Physical Review Research}\ }\textbf {\bibinfo {volume}
  {6}},\ \bibinfo {pages} {023228} (\bibinfo {year} {2024})}\BibitemShut
  {NoStop}%
\bibitem [{\citenamefont {Papo}\ and\ \citenamefont {Buldú}(2024)}]{Papo2024}%
  \BibitemOpen
  \bibfield  {author} {\bibinfo {author} {\bibfnamefont {D.}~\bibnamefont
  {Papo}}\ and\ \bibinfo {author} {\bibfnamefont {J.}~\bibnamefont {Buldú}},\
  }\bibfield  {title} {\bibinfo {title} {Does the brain behave like a (complex)
  network? i. dynamics},\ }\href
  {https://doi.org/https://doi.org/10.1016/j.plrev.2023.12.006} {\bibfield
  {journal} {\bibinfo  {journal} {Physics of Life Reviews}\ }\textbf {\bibinfo
  {volume} {48}},\ \bibinfo {pages} {47} (\bibinfo {year} {2024})}\BibitemShut
  {NoStop}%
\bibitem [{\citenamefont {Motter}\ \emph {et~al.}(2013)\citenamefont {Motter},
  \citenamefont {Myers}, \citenamefont {Anghel},\ and\ \citenamefont
  {Nishikawa}}]{Motter2013}%
  \BibitemOpen
  \bibfield  {author} {\bibinfo {author} {\bibfnamefont {A.~E.}\ \bibnamefont
  {Motter}}, \bibinfo {author} {\bibfnamefont {S.~A.}\ \bibnamefont {Myers}},
  \bibinfo {author} {\bibfnamefont {M.}~\bibnamefont {Anghel}},\ and\ \bibinfo
  {author} {\bibfnamefont {T.}~\bibnamefont {Nishikawa}},\ }\bibfield  {title}
  {\bibinfo {title} {Spontaneous synchrony in power-grid networks},\ }\href
  {https://doi.org/10.1038/nphys2535} {\bibfield  {journal} {\bibinfo
  {journal} {Nature Physics}\ }\textbf {\bibinfo {volume} {9}},\ \bibinfo
  {pages} {191} (\bibinfo {year} {2013})}\BibitemShut {NoStop}%
\bibitem [{\citenamefont {Aparicio}\ \emph {et~al.}(2021)\citenamefont
  {Aparicio}, \citenamefont {Velasco-Hernández}, \citenamefont {Moog},
  \citenamefont {Liu},\ and\ \citenamefont {Angulo}}]{Aparicio2021}%
  \BibitemOpen
  \bibfield  {author} {\bibinfo {author} {\bibfnamefont {A.}~\bibnamefont
  {Aparicio}}, \bibinfo {author} {\bibfnamefont {J.~X.}\ \bibnamefont
  {Velasco-Hernández}}, \bibinfo {author} {\bibfnamefont {C.~H.}\ \bibnamefont
  {Moog}}, \bibinfo {author} {\bibfnamefont {Y.-Y.}\ \bibnamefont {Liu}},\ and\
  \bibinfo {author} {\bibfnamefont {M.~T.}\ \bibnamefont {Angulo}},\ }\bibfield
   {title} {\bibinfo {title} {Structure-based identification of sensor species
  for anticipating critical transitions},\ }\href
  {https://doi.org/10.1073/pnas.2104732118} {\bibfield  {journal} {\bibinfo
  {journal} {Proceedings of the National Academy of Sciences}\ }\textbf
  {\bibinfo {volume} {118}},\ \bibinfo {pages} {e2104732118} (\bibinfo {year}
  {2021})}\BibitemShut {NoStop}%
\bibitem [{\citenamefont {Fan}\ \emph {et~al.}(2021)\citenamefont {Fan},
  \citenamefont {Kong}, \citenamefont {Lai},\ and\ \citenamefont
  {Wang}}]{Fan2021}%
  \BibitemOpen
  \bibfield  {author} {\bibinfo {author} {\bibfnamefont {H.}~\bibnamefont
  {Fan}}, \bibinfo {author} {\bibfnamefont {L.~W.}\ \bibnamefont {Kong}},
  \bibinfo {author} {\bibfnamefont {Y.~C.}\ \bibnamefont {Lai}},\ and\ \bibinfo
  {author} {\bibfnamefont {X.}~\bibnamefont {Wang}},\ }\bibfield  {title}
  {\bibinfo {title} {Anticipating synchronization with machine learning},\
  }\href {https://doi.org/10.1103/PhysRevResearch.3.023237} {\bibfield
  {journal} {\bibinfo  {journal} {Physical Review Research}\ }\textbf {\bibinfo
  {volume} {3}},\ \bibinfo {pages} {023237} (\bibinfo {year}
  {2021})}\BibitemShut {NoStop}%
\bibitem [{\citenamefont {Masuda}\ \emph {et~al.}(2024)\citenamefont {Masuda},
  \citenamefont {Aihara},\ and\ \citenamefont {MacLaren}}]{Masuda2024}%
  \BibitemOpen
  \bibfield  {author} {\bibinfo {author} {\bibfnamefont {N.}~\bibnamefont
  {Masuda}}, \bibinfo {author} {\bibfnamefont {K.}~\bibnamefont {Aihara}},\
  and\ \bibinfo {author} {\bibfnamefont {N.~G.}\ \bibnamefont {MacLaren}},\
  }\bibfield  {title} {\bibinfo {title} {Anticipating regime shifts by mixing
  early warning signals from different nodes},\ }\href
  {https://doi.org/10.1038/s41467-024-45476-9} {\bibfield  {journal} {\bibinfo
  {journal} {Nature Communications}\ }\textbf {\bibinfo {volume} {15}},\
  \bibinfo {pages} {1086} (\bibinfo {year} {2024})}\BibitemShut {NoStop}%
\bibitem [{\citenamefont {MacLaren}\ \emph {et~al.}(2023)\citenamefont
  {MacLaren}, \citenamefont {Kundu},\ and\ \citenamefont
  {Masuda}}]{MacLaren2023}%
  \BibitemOpen
  \bibfield  {author} {\bibinfo {author} {\bibfnamefont {N.~G.}\ \bibnamefont
  {MacLaren}}, \bibinfo {author} {\bibfnamefont {P.}~\bibnamefont {Kundu}},\
  and\ \bibinfo {author} {\bibfnamefont {N.}~\bibnamefont {Masuda}},\
  }\bibfield  {title} {\bibinfo {title} {Early warnings for multi-stage
  transitions in dynamics on networks},\ }\href
  {https://doi.org/10.1098/rsif.2022.0743} {\bibfield  {journal} {\bibinfo
  {journal} {Journal of the Royal Society Interface}\ }\textbf {\bibinfo
  {volume} {20}},\ \bibinfo {pages} {20220743} (\bibinfo {year}
  {2023})}\BibitemShut {NoStop}%
\bibitem [{\citenamefont {Ehstand}\ \emph {et~al.}(2023)\citenamefont
  {Ehstand}, \citenamefont {Donner}, \citenamefont {López},\ and\
  \citenamefont {Hernández-García}}]{Ehstand2023}%
  \BibitemOpen
  \bibfield  {author} {\bibinfo {author} {\bibfnamefont {N.}~\bibnamefont
  {Ehstand}}, \bibinfo {author} {\bibfnamefont {R.~V.}\ \bibnamefont {Donner}},
  \bibinfo {author} {\bibfnamefont {C.}~\bibnamefont {López}},\ and\ \bibinfo
  {author} {\bibfnamefont {E.}~\bibnamefont {Hernández-García}},\ }\bibfield
  {title} {\bibinfo {title} {Network percolation provides early warnings of
  abrupt changes in coupled oscillatory systems: An explanatory analysis},\
  }\href {https://doi.org/10.1103/PhysRevE.108.054207} {\bibfield  {journal}
  {\bibinfo  {journal} {Physical Review E}\ }\textbf {\bibinfo {volume}
  {108}},\ \bibinfo {pages} {054207} (\bibinfo {year} {2023})}\BibitemShut
  {NoStop}%
\bibitem [{\citenamefont {MacLaren}\ \emph {et~al.}(2024)\citenamefont
  {MacLaren}, \citenamefont {Aihara},\ and\ \citenamefont
  {Masuda}}]{MacLaren2024a}%
  \BibitemOpen
  \bibfield  {author} {\bibinfo {author} {\bibfnamefont {N.~G.}\ \bibnamefont
  {MacLaren}}, \bibinfo {author} {\bibfnamefont {K.}~\bibnamefont {Aihara}},\
  and\ \bibinfo {author} {\bibfnamefont {N.}~\bibnamefont {Masuda}},\ }\href
  {https://doi.org/10.48550/arXiv.2410.04303} {\bibinfo {title} {Applicability
  of spatial early warning signals to complex network dynamics}} (\bibinfo
  {year} {2024})\BibitemShut {NoStop}%
\bibitem [{\citenamefont {Bury}\ \emph {et~al.}(2021)\citenamefont {Bury},
  \citenamefont {Sujith}, \citenamefont {Pavithran}, \citenamefont {Scheffer},
  \citenamefont {Lenton}, \citenamefont {Anand},\ and\ \citenamefont
  {Bauch}}]{Bury2021}%
  \BibitemOpen
  \bibfield  {author} {\bibinfo {author} {\bibfnamefont {T.~M.}\ \bibnamefont
  {Bury}}, \bibinfo {author} {\bibfnamefont {R.~I.}\ \bibnamefont {Sujith}},
  \bibinfo {author} {\bibfnamefont {I.}~\bibnamefont {Pavithran}}, \bibinfo
  {author} {\bibfnamefont {M.}~\bibnamefont {Scheffer}}, \bibinfo {author}
  {\bibfnamefont {T.~M.}\ \bibnamefont {Lenton}}, \bibinfo {author}
  {\bibfnamefont {M.}~\bibnamefont {Anand}},\ and\ \bibinfo {author}
  {\bibfnamefont {C.~T.}\ \bibnamefont {Bauch}},\ }\bibfield  {title} {\bibinfo
  {title} {Deep learning for early warning signals of tipping points},\ }\href
  {https://doi.org/10.1073/pnas.2106140118} {\bibfield  {journal} {\bibinfo
  {journal} {Proceedings of the National Academy of Sciences}\ }\textbf
  {\bibinfo {volume} {118}},\ \bibinfo {pages} {e2106140118} (\bibinfo {year}
  {2021})}\BibitemShut {NoStop}%
\bibitem [{\citenamefont {Kong}\ \emph {et~al.}(2021)\citenamefont {Kong},
  \citenamefont {Fan}, \citenamefont {Grebogi},\ and\ \citenamefont
  {Lai}}]{Kong2021}%
  \BibitemOpen
  \bibfield  {author} {\bibinfo {author} {\bibfnamefont {L.~W.}\ \bibnamefont
  {Kong}}, \bibinfo {author} {\bibfnamefont {H.~W.}\ \bibnamefont {Fan}},
  \bibinfo {author} {\bibfnamefont {C.}~\bibnamefont {Grebogi}},\ and\ \bibinfo
  {author} {\bibfnamefont {Y.~C.}\ \bibnamefont {Lai}},\ }\bibfield  {title}
  {\bibinfo {title} {Machine learning prediction of critical transition and
  system collapse},\ }\href {https://doi.org/10.1103/PhysRevResearch.3.013090}
  {\bibfield  {journal} {\bibinfo  {journal} {Physical Review Research}\
  }\textbf {\bibinfo {volume} {3}},\ \bibinfo {pages} {013090} (\bibinfo {year}
  {2021})}\BibitemShut {NoStop}%
\bibitem [{\citenamefont {Vishnoi}\ \emph {et~al.}(2024)\citenamefont
  {Vishnoi}, \citenamefont {Gupta}, \citenamefont {Saurabh},\ and\
  \citenamefont {Kabiraj}}]{Vishnoi2024}%
  \BibitemOpen
  \bibfield  {author} {\bibinfo {author} {\bibfnamefont {N.}~\bibnamefont
  {Vishnoi}}, \bibinfo {author} {\bibfnamefont {V.}~\bibnamefont {Gupta}},
  \bibinfo {author} {\bibfnamefont {A.}~\bibnamefont {Saurabh}},\ and\ \bibinfo
  {author} {\bibfnamefont {L.}~\bibnamefont {Kabiraj}},\ }\bibfield  {title}
  {\bibinfo {title} {Reliability of early warning indicators of critical
  transition in stochastic van der pol oscillators with additive correlated
  noise},\ }\href {https://doi.org/10.1007/s11071-024-09831-1} {\bibfield
  {journal} {\bibinfo  {journal} {Nonlinear Dynamics}\ }\textbf {\bibinfo
  {volume} {112}},\ \bibinfo {pages} {15193} (\bibinfo {year}
  {2024})}\BibitemShut {NoStop}%
\bibitem [{\citenamefont {Tarigo}\ \emph {et~al.}(2024)\citenamefont {Tarigo},
  \citenamefont {Stari}, \citenamefont {Masoller},\ and\ \citenamefont
  {Martí}}]{Tarigo2024}%
  \BibitemOpen
  \bibfield  {author} {\bibinfo {author} {\bibfnamefont {J.~P.}\ \bibnamefont
  {Tarigo}}, \bibinfo {author} {\bibfnamefont {C.}~\bibnamefont {Stari}},
  \bibinfo {author} {\bibfnamefont {C.}~\bibnamefont {Masoller}},\ and\
  \bibinfo {author} {\bibfnamefont {A.~C.}\ \bibnamefont {Martí}},\ }\bibfield
   {title} {\bibinfo {title} {Basin entropy as an indicator of a bifurcation in
  a time-delayed system},\ }\href {https://doi.org/10.1063/5.0201932}
  {\bibfield  {journal} {\bibinfo  {journal} {Chaos}\ }\textbf {\bibinfo
  {volume} {34}},\ \bibinfo {pages} {053113} (\bibinfo {year}
  {2024})}\BibitemShut {NoStop}%
\bibitem [{\citenamefont {Bassi}\ \emph {et~al.}(2022)\citenamefont {Bassi},
  \citenamefont {Yim}, \citenamefont {Vendrow}, \citenamefont {Koduluka},
  \citenamefont {Zhu},\ and\ \citenamefont {Lyu}}]{Bassi2022}%
  \BibitemOpen
  \bibfield  {author} {\bibinfo {author} {\bibfnamefont {H.}~\bibnamefont
  {Bassi}}, \bibinfo {author} {\bibfnamefont {R.~P.}\ \bibnamefont {Yim}},
  \bibinfo {author} {\bibfnamefont {J.}~\bibnamefont {Vendrow}}, \bibinfo
  {author} {\bibfnamefont {R.}~\bibnamefont {Koduluka}}, \bibinfo {author}
  {\bibfnamefont {C.}~\bibnamefont {Zhu}},\ and\ \bibinfo {author}
  {\bibfnamefont {H.}~\bibnamefont {Lyu}},\ }\bibfield  {title} {\bibinfo
  {title} {Learning to predict synchronization of coupled oscillators on
  randomly generated graphs},\ }\href
  {https://doi.org/10.1038/s41598-022-18953-8} {\bibfield  {journal} {\bibinfo
  {journal} {Scientific Reports}\ }\textbf {\bibinfo {volume} {12}},\ \bibinfo
  {pages} {15056} (\bibinfo {year} {2022})}\BibitemShut {NoStop}%
\bibitem [{\citenamefont {Ma}\ \emph {et~al.}(2023)\citenamefont {Ma},
  \citenamefont {Dai}, \citenamefont {Li},\ and\ \citenamefont
  {Yang}}]{Ma2023}%
  \BibitemOpen
  \bibfield  {author} {\bibinfo {author} {\bibfnamefont {R.}~\bibnamefont
  {Ma}}, \bibinfo {author} {\bibfnamefont {Q.}~\bibnamefont {Dai}}, \bibinfo
  {author} {\bibfnamefont {H.}~\bibnamefont {Li}},\ and\ \bibinfo {author}
  {\bibfnamefont {J.}~\bibnamefont {Yang}},\ }\bibfield  {title} {\bibinfo
  {title} {Dynamics reconstruction in the presence of bistability by using
  reservoir computer},\ }\href {https://doi.org/10.1016/j.chaos.2023.113523}
  {\bibfield  {journal} {\bibinfo  {journal} {Chaos, Solitons \& Fractals}\
  }\textbf {\bibinfo {volume} {172}},\ \bibinfo {pages} {113523} (\bibinfo
  {year} {2023})}\BibitemShut {NoStop}%
\bibitem [{\citenamefont {Zhang}\ \emph {et~al.}(2022)\citenamefont {Zhang},
  \citenamefont {Fan}, \citenamefont {Du}, \citenamefont {Wang},\ and\
  \citenamefont {Wang}}]{Zhang2022}%
  \BibitemOpen
  \bibfield  {author} {\bibinfo {author} {\bibfnamefont {H.}~\bibnamefont
  {Zhang}}, \bibinfo {author} {\bibfnamefont {H.}~\bibnamefont {Fan}}, \bibinfo
  {author} {\bibfnamefont {Y.}~\bibnamefont {Du}}, \bibinfo {author}
  {\bibfnamefont {L.}~\bibnamefont {Wang}},\ and\ \bibinfo {author}
  {\bibfnamefont {X.}~\bibnamefont {Wang}},\ }\bibfield  {title} {\bibinfo
  {title} {Anticipating measure synchronization in coupled hamiltonian systems
  with machine learning},\ }\href {https://doi.org/10.1063/5.0093663}
  {\bibfield  {journal} {\bibinfo  {journal} {Chaos}\ }\textbf {\bibinfo
  {volume} {32}},\ \bibinfo {pages} {083136} (\bibinfo {year}
  {2022})}\BibitemShut {NoStop}%
\bibitem [{\citenamefont {Liu}\ \emph {et~al.}(2024)\citenamefont {Liu},
  \citenamefont {Zhang}, \citenamefont {Ru}, \citenamefont {Gao}, \citenamefont
  {Moore},\ and\ \citenamefont {Yan}}]{Liu2024}%
  \BibitemOpen
  \bibfield  {author} {\bibinfo {author} {\bibfnamefont {Z.}~\bibnamefont
  {Liu}}, \bibinfo {author} {\bibfnamefont {X.}~\bibnamefont {Zhang}}, \bibinfo
  {author} {\bibfnamefont {X.}~\bibnamefont {Ru}}, \bibinfo {author}
  {\bibfnamefont {T.-T.}\ \bibnamefont {Gao}}, \bibinfo {author} {\bibfnamefont
  {J.~M.}\ \bibnamefont {Moore}},\ and\ \bibinfo {author} {\bibfnamefont
  {G.}~\bibnamefont {Yan}},\ }\bibfield  {title} {\bibinfo {title} {Early
  predictor for the onset of critical transitions in networked dynamical
  systems},\ }\href {https://doi.org/10.1103/PhysRevX.14.031009} {\bibfield
  {journal} {\bibinfo  {journal} {Physical Review X}\ }\textbf {\bibinfo
  {volume} {14}},\ \bibinfo {pages} {031009} (\bibinfo {year}
  {2024})}\BibitemShut {NoStop}%
\bibitem [{\citenamefont {Roy}\ \emph {et~al.}(2022)\citenamefont {Roy},
  \citenamefont {Senapati}, \citenamefont {Poria}, \citenamefont {Mishra},\
  and\ \citenamefont {Hens}}]{Roy2022}%
  \BibitemOpen
  \bibfield  {author} {\bibinfo {author} {\bibfnamefont {M.}~\bibnamefont
  {Roy}}, \bibinfo {author} {\bibfnamefont {A.}~\bibnamefont {Senapati}},
  \bibinfo {author} {\bibfnamefont {S.}~\bibnamefont {Poria}}, \bibinfo
  {author} {\bibfnamefont {A.}~\bibnamefont {Mishra}},\ and\ \bibinfo {author}
  {\bibfnamefont {C.}~\bibnamefont {Hens}},\ }\bibfield  {title} {\bibinfo
  {title} {Role of assortativity in predicting burst synchronization using echo
  state network},\ }\href {https://doi.org/10.1103/PhysRevE.105.064205}
  {\bibfield  {journal} {\bibinfo  {journal} {Physical Review E}\ }\textbf
  {\bibinfo {volume} {105}},\ \bibinfo {pages} {064205} (\bibinfo {year}
  {2022})}\BibitemShut {NoStop}%
\bibitem [{\citenamefont {Shahriari}\ \emph {et~al.}(2023)\citenamefont
  {Shahriari}, \citenamefont {Algar}, \citenamefont {Walker},\ and\
  \citenamefont {Small}}]{Shahriari2023}%
  \BibitemOpen
  \bibfield  {author} {\bibinfo {author} {\bibfnamefont {Z.}~\bibnamefont
  {Shahriari}}, \bibinfo {author} {\bibfnamefont {S.~D.}\ \bibnamefont
  {Algar}}, \bibinfo {author} {\bibfnamefont {D.~M.}\ \bibnamefont {Walker}},\
  and\ \bibinfo {author} {\bibfnamefont {M.}~\bibnamefont {Small}},\ }\bibfield
   {title} {\bibinfo {title} {Ordinal poincar{\'e} sections: Reconstructing the
  first return map from an ordinal segmentation of time series},\ }\href
  {https://doi.org/10.1063/5.0141438} {\bibfield  {journal} {\bibinfo
  {journal} {Chaos}\ }\textbf {\bibinfo {volume} {33}},\ \bibinfo {pages}
  {053109} (\bibinfo {year} {2023})}\BibitemShut {NoStop}%
\bibitem [{\citenamefont {Leyva}\ and\ \citenamefont
  {Masoller}(2020)}]{Leyva2020}%
  \BibitemOpen
  \bibfield  {author} {\bibinfo {author} {\bibfnamefont {I.}~\bibnamefont
  {Leyva}}\ and\ \bibinfo {author} {\bibfnamefont {C.}~\bibnamefont
  {Masoller}},\ }\bibfield  {title} {\bibinfo {title} {Inferring the
  connectivity of coupled oscillators and anticipating their transition to
  synchrony through lag-time analysis},\ }\href
  {https://doi.org/10.1016/j.chaos.2020.109604} {\bibfield  {journal} {\bibinfo
   {journal} {Chaos, Solitons \& Fractals}\ }\textbf {\bibinfo {volume}
  {133}},\ \bibinfo {pages} {109604} (\bibinfo {year} {2020})}\BibitemShut
  {NoStop}%
\bibitem [{\citenamefont {Leyva}\ \emph {et~al.}(2022)\citenamefont {Leyva},
  \citenamefont {Mart{\'\i}nez}, \citenamefont {Masoller}, \citenamefont
  {Rosso},\ and\ \citenamefont {Zanin}}]{Leyva2022}%
  \BibitemOpen
  \bibfield  {author} {\bibinfo {author} {\bibfnamefont {I.}~\bibnamefont
  {Leyva}}, \bibinfo {author} {\bibfnamefont {J.~H.}\ \bibnamefont
  {Mart{\'\i}nez}}, \bibinfo {author} {\bibfnamefont {C.}~\bibnamefont
  {Masoller}}, \bibinfo {author} {\bibfnamefont {O.~A.}\ \bibnamefont
  {Rosso}},\ and\ \bibinfo {author} {\bibfnamefont {M.}~\bibnamefont {Zanin}},\
  }\bibfield  {title} {\bibinfo {title} {20 years of ordinal patterns:
  Perspectives and challenges},\ }\href
  {https://doi.org/10.1209/0295-5075/ac6a72} {\bibfield  {journal} {\bibinfo
  {journal} {Europhysics Letters}\ }\textbf {\bibinfo {volume} {138}},\
  \bibinfo {pages} {31001} (\bibinfo {year} {2022})}\BibitemShut {NoStop}%
\bibitem [{\citenamefont {Almendral}\ \emph {et~al.}(2023)\citenamefont
  {Almendral}, \citenamefont {Leyva},\ and\ \citenamefont
  {Sendiña-Nadal}}]{Almendral2023}%
  \BibitemOpen
  \bibfield  {author} {\bibinfo {author} {\bibfnamefont {J.~A.}\ \bibnamefont
  {Almendral}}, \bibinfo {author} {\bibfnamefont {I.}~\bibnamefont {Leyva}},\
  and\ \bibinfo {author} {\bibfnamefont {I.}~\bibnamefont {Sendiña-Nadal}},\
  }\bibfield  {title} {\bibinfo {title} {Unveiling the connectivity of complex
  networks using ordinal transition methods},\ }\href
  {https://doi.org/10.3390/e25071079} {\bibfield  {journal} {\bibinfo
  {journal} {Entropy}\ }\textbf {\bibinfo {volume} {25}},\ \bibinfo {pages}
  {1079} (\bibinfo {year} {2023})}\BibitemShut {NoStop}%
\bibitem [{\citenamefont {Tirabassi}\ and\ \citenamefont
  {Masoller}(2023)}]{Tirabassi2023}%
  \BibitemOpen
  \bibfield  {author} {\bibinfo {author} {\bibfnamefont {G.}~\bibnamefont
  {Tirabassi}}\ and\ \bibinfo {author} {\bibfnamefont {C.}~\bibnamefont
  {Masoller}},\ }\bibfield  {title} {\bibinfo {title} {Entropy-based early
  detection of critical transitions in spatial vegetation fields},\ }\href
  {https://doi.org/10.1073/pnas.2215667120} {\bibfield  {journal} {\bibinfo
  {journal} {Proceedings of the National Academy of Sciences}\ }\textbf
  {\bibinfo {volume} {120}},\ \bibinfo {pages} {e2215667120} (\bibinfo {year}
  {2023})}\BibitemShut {NoStop}%
\bibitem [{\citenamefont {Lehnertz}(2023)}]{Lehnertz2023}%
  \BibitemOpen
  \bibfield  {author} {\bibinfo {author} {\bibfnamefont {K.}~\bibnamefont
  {Lehnertz}},\ }\bibfield  {title} {\bibinfo {title} {{Ordinal methods for a
  characterization of evolving functional brain networks}},\ }\href
  {https://doi.org/10.1063/5.0136181} {\bibfield  {journal} {\bibinfo
  {journal} {Chaos}\ }\textbf {\bibinfo {volume} {33}},\ \bibinfo {pages}
  {022101} (\bibinfo {year} {2023})}\BibitemShut {NoStop}%
\bibitem [{\citenamefont {Boccaletti}\ \emph {et~al.}(2016)\citenamefont
  {Boccaletti}, \citenamefont {Almendral}, \citenamefont {Guan}, \citenamefont
  {Leyva}, \citenamefont {Liu}, \citenamefont {Sendiña-Nadal}, \citenamefont
  {Wang},\ and\ \citenamefont {Zou}}]{Boccaletti2016}%
  \BibitemOpen
  \bibfield  {author} {\bibinfo {author} {\bibfnamefont {S.}~\bibnamefont
  {Boccaletti}}, \bibinfo {author} {\bibfnamefont {J.}~\bibnamefont
  {Almendral}}, \bibinfo {author} {\bibfnamefont {S.}~\bibnamefont {Guan}},
  \bibinfo {author} {\bibfnamefont {I.}~\bibnamefont {Leyva}}, \bibinfo
  {author} {\bibfnamefont {Z.}~\bibnamefont {Liu}}, \bibinfo {author}
  {\bibfnamefont {I.}~\bibnamefont {Sendiña-Nadal}}, \bibinfo {author}
  {\bibfnamefont {Z.}~\bibnamefont {Wang}},\ and\ \bibinfo {author}
  {\bibfnamefont {Y.}~\bibnamefont {Zou}},\ }\bibfield  {title} {\bibinfo
  {title} {Explosive transitions in complex networks’ structure and dynamics:
  Percolation and synchronization},\ }\href
  {https://doi.org/10.1016/j.physrep.2016.10.004} {\bibfield  {journal}
  {\bibinfo  {journal} {Physics Reports}\ }\textbf {\bibinfo {volume} {660}},\
  \bibinfo {pages} {1} (\bibinfo {year} {2016})}\BibitemShut {NoStop}%
\bibitem [{\citenamefont {Avalos-Gayt\'an}\ \emph {et~al.}(2018)\citenamefont
  {Avalos-Gayt\'an}, \citenamefont {Almendral}, \citenamefont {Leyva},
  \citenamefont {Battiston}, \citenamefont {Nicosia}, \citenamefont {Latora},\
  and\ \citenamefont {Boccaletti}}]{Avalos2018}%
  \BibitemOpen
  \bibfield  {author} {\bibinfo {author} {\bibfnamefont {V.}~\bibnamefont
  {Avalos-Gayt\'an}}, \bibinfo {author} {\bibfnamefont {J.~A.}\ \bibnamefont
  {Almendral}}, \bibinfo {author} {\bibfnamefont {I.}~\bibnamefont {Leyva}},
  \bibinfo {author} {\bibfnamefont {F.}~\bibnamefont {Battiston}}, \bibinfo
  {author} {\bibfnamefont {V.}~\bibnamefont {Nicosia}}, \bibinfo {author}
  {\bibfnamefont {V.}~\bibnamefont {Latora}},\ and\ \bibinfo {author}
  {\bibfnamefont {S.}~\bibnamefont {Boccaletti}},\ }\bibfield  {title}
  {\bibinfo {title} {Emergent explosive synchronization in adaptive complex
  networks},\ }\href {https://doi.org/10.1103/PhysRevE.97.042301} {\bibfield
  {journal} {\bibinfo  {journal} {Physical Review E}\ }\textbf {\bibinfo
  {volume} {97}},\ \bibinfo {pages} {042301} (\bibinfo {year}
  {2018})}\BibitemShut {NoStop}%
\bibitem [{\citenamefont {Soriano-Pa{\~n}os}\ \emph {et~al.}(2019)\citenamefont
  {Soriano-Pa{\~n}os}, \citenamefont {Guo}, \citenamefont {Latora},\ and\
  \citenamefont {G{\'o}mez-Garde{\~n}es}}]{Soriano2019}%
  \BibitemOpen
  \bibfield  {author} {\bibinfo {author} {\bibfnamefont {D.}~\bibnamefont
  {Soriano-Pa{\~n}os}}, \bibinfo {author} {\bibfnamefont {Q.}~\bibnamefont
  {Guo}}, \bibinfo {author} {\bibfnamefont {V.}~\bibnamefont {Latora}},\ and\
  \bibinfo {author} {\bibfnamefont {J.}~\bibnamefont
  {G{\'o}mez-Garde{\~n}es}},\ }\bibfield  {title} {\bibinfo {title} {Explosive
  transitions induced by interdependent contagion-consensus dynamics in
  multiplex networks},\ }\href {https://doi.org/10.1103/PhysRevE.99.062311}
  {\bibfield  {journal} {\bibinfo  {journal} {Physical Review E}\ }\textbf
  {\bibinfo {volume} {99}},\ \bibinfo {pages} {062311} (\bibinfo {year}
  {2019})}\BibitemShut {NoStop}%
\bibitem [{\citenamefont {Khalil}\ \emph {et~al.}(2023)\citenamefont {Khalil},
  \citenamefont {Leyva}, \citenamefont {Almendral},\ and\ \citenamefont
  {Sendi\~na Nadal}}]{Khalil2023}%
  \BibitemOpen
  \bibfield  {author} {\bibinfo {author} {\bibfnamefont {N.}~\bibnamefont
  {Khalil}}, \bibinfo {author} {\bibfnamefont {I.}~\bibnamefont {Leyva}},
  \bibinfo {author} {\bibfnamefont {J.~A.}\ \bibnamefont {Almendral}},\ and\
  \bibinfo {author} {\bibfnamefont {I.}~\bibnamefont {Sendi\~na Nadal}},\
  }\bibfield  {title} {\bibinfo {title} {Deterministic and stochastic
  cooperation transitions in evolutionary games on networks},\ }\href
  {https://doi.org/10.1103/PhysRevE.107.054302} {\bibfield  {journal} {\bibinfo
   {journal} {Physical Review E}\ }\textbf {\bibinfo {volume} {107}},\ \bibinfo
  {pages} {054302} (\bibinfo {year} {2023})}\BibitemShut {NoStop}%
\bibitem [{\citenamefont {Ranjan}\ and\ \citenamefont
  {Gandhi}(2024)}]{Ranjan2024}%
  \BibitemOpen
  \bibfield  {author} {\bibinfo {author} {\bibfnamefont {A.}~\bibnamefont
  {Ranjan}}\ and\ \bibinfo {author} {\bibfnamefont {S.~R.}\ \bibnamefont
  {Gandhi}},\ }\bibfield  {title} {\bibinfo {title} {Propagation of transient
  explosive synchronization in a mesoscale mouse brain network model of
  epilepsy},\ }\href {https://doi.org/10.1162/netn_a_00379} {\bibfield
  {journal} {\bibinfo  {journal} {Network Neuroscience}\ }\textbf {\bibinfo
  {volume} {8}},\ \bibinfo {pages} {883} (\bibinfo {year} {2024})}\BibitemShut
  {NoStop}%
\bibitem [{\citenamefont {Rahjerdi}\ \emph {et~al.}(2022)\citenamefont
  {Rahjerdi}, \citenamefont {Ramamoorthy}, \citenamefont {Nazarimehr},
  \citenamefont {Rajagopal},\ and\ \citenamefont {Jafari}}]{Karimi2022}%
  \BibitemOpen
  \bibfield  {author} {\bibinfo {author} {\bibfnamefont {B.~K.}\ \bibnamefont
  {Rahjerdi}}, \bibinfo {author} {\bibfnamefont {R.}~\bibnamefont
  {Ramamoorthy}}, \bibinfo {author} {\bibfnamefont {F.}~\bibnamefont
  {Nazarimehr}}, \bibinfo {author} {\bibfnamefont {K.}~\bibnamefont
  {Rajagopal}},\ and\ \bibinfo {author} {\bibfnamefont {S.}~\bibnamefont
  {Jafari}},\ }\bibfield  {title} {\bibinfo {title} {Indicating the
  synchronization bifurcation points using the early warning signals in two
  case studies: Continuous and explosive synchronization},\ }\href
  {https://doi.org/10.1016/j.chaos.2022.112656} {\bibfield  {journal} {\bibinfo
   {journal} {Chaos, Solitons \& Fractals}\ }\textbf {\bibinfo {volume}
  {164}},\ \bibinfo {pages} {112656} (\bibinfo {year} {2022})}\BibitemShut
  {NoStop}%
\bibitem [{\citenamefont {Tlaie}\ \emph {et~al.}(2019)\citenamefont {Tlaie},
  \citenamefont {Leyva}, \citenamefont {Sevilla-Escoboza}, \citenamefont
  {Vera-Avila},\ and\ \citenamefont {Sendi\~na Nadal}}]{Tlaie2019}%
  \BibitemOpen
  \bibfield  {author} {\bibinfo {author} {\bibfnamefont {A.}~\bibnamefont
  {Tlaie}}, \bibinfo {author} {\bibfnamefont {I.}~\bibnamefont {Leyva}},
  \bibinfo {author} {\bibfnamefont {R.}~\bibnamefont {Sevilla-Escoboza}},
  \bibinfo {author} {\bibfnamefont {V.~P.}\ \bibnamefont {Vera-Avila}},\ and\
  \bibinfo {author} {\bibfnamefont {I.}~\bibnamefont {Sendi\~na Nadal}},\
  }\bibfield  {title} {\bibinfo {title} {Dynamical complexity as a proxy for
  the network degree distribution},\ }\href
  {https://doi.org/10.1103/PhysRevE.99.012310} {\bibfield  {journal} {\bibinfo
  {journal} {Physical Review E}\ }\textbf {\bibinfo {volume} {99}},\ \bibinfo
  {pages} {012310} (\bibinfo {year} {2019})}\BibitemShut {NoStop}%
\bibitem [{\citenamefont {Bandt}\ and\ \citenamefont
  {Pompe}(2002)}]{Bandt2002}%
  \BibitemOpen
  \bibfield  {author} {\bibinfo {author} {\bibfnamefont {C.}~\bibnamefont
  {Bandt}}\ and\ \bibinfo {author} {\bibfnamefont {B.}~\bibnamefont {Pompe}},\
  }\bibfield  {title} {\bibinfo {title} {Permutation entropy: A natural
  complexity measure for time series},\ }\href
  {https://doi.org/10.1103/PhysRevLett.88.174102} {\bibfield  {journal}
  {\bibinfo  {journal} {Physical Review Letters}\ }\textbf {\bibinfo {volume}
  {88}},\ \bibinfo {pages} {174102} (\bibinfo {year} {2002})}\BibitemShut
  {NoStop}%
\bibitem [{\citenamefont {McCullough}\ \emph {et~al.}(2015)\citenamefont
  {McCullough}, \citenamefont {Small}, \citenamefont {Stemler}, \citenamefont
  {Ho}, \citenamefont {Iu},\ and\ \citenamefont {Iu}}]{McCullough2015}%
  \BibitemOpen
  \bibfield  {author} {\bibinfo {author} {\bibfnamefont {M.}~\bibnamefont
  {McCullough}}, \bibinfo {author} {\bibfnamefont {M.}~\bibnamefont {Small}},
  \bibinfo {author} {\bibfnamefont {T.}~\bibnamefont {Stemler}}, \bibinfo
  {author} {\bibfnamefont {H.}~\bibnamefont {Ho}}, \bibinfo {author}
  {\bibfnamefont {C.}~\bibnamefont {Iu}},\ and\ \bibinfo {author}
  {\bibfnamefont {H.~H.~C.}\ \bibnamefont {Iu}},\ }\bibfield  {title} {\bibinfo
  {title} {{Time lagged ordinal partition networks for capturing dynamics of
  continuous dynamical systems}},\ }\href {https://doi.org/10.1063/1.4919075}
  {\bibfield  {journal} {\bibinfo  {journal} {Chaos}\ }\textbf {\bibinfo
  {volume} {25}},\ \bibinfo {pages} {53101} (\bibinfo {year}
  {2015})}\BibitemShut {NoStop}%
\bibitem [{\citenamefont {Zanin}\ and\ \citenamefont
  {Olivares}(2021)}]{Zanin21}%
  \BibitemOpen
  \bibfield  {author} {\bibinfo {author} {\bibfnamefont {M.}~\bibnamefont
  {Zanin}}\ and\ \bibinfo {author} {\bibfnamefont {F.}~\bibnamefont
  {Olivares}},\ }\bibfield  {title} {\bibinfo {title} {Ordinal patterns-based
  methodologies for distinguishing chaos from noise in discrete time series},\
  }\href {https://doi.org/10.1038/s42005-021-00696-z} {\bibfield  {journal}
  {\bibinfo  {journal} {Communications Physics}\ }\textbf {\bibinfo {volume}
  {4}},\ \bibinfo {pages} {190} (\bibinfo {year} {2021})}\BibitemShut {NoStop}%
\bibitem [{\citenamefont {Zunino}\ \emph {et~al.}(2010)\citenamefont {Zunino},
  \citenamefont {Soriano}, \citenamefont {Fischer}, \citenamefont {Rosso},\
  and\ \citenamefont {Mirasso}}]{zunino2010}%
  \BibitemOpen
  \bibfield  {author} {\bibinfo {author} {\bibfnamefont {L.}~\bibnamefont
  {Zunino}}, \bibinfo {author} {\bibfnamefont {M.~C.}\ \bibnamefont {Soriano}},
  \bibinfo {author} {\bibfnamefont {I.}~\bibnamefont {Fischer}}, \bibinfo
  {author} {\bibfnamefont {O.~A.}\ \bibnamefont {Rosso}},\ and\ \bibinfo
  {author} {\bibfnamefont {C.~R.}\ \bibnamefont {Mirasso}},\ }\bibfield
  {title} {\bibinfo {title} {Permutation-information-theory approach to unveil
  delay dynamics from time-series analysis},\ }\href
  {https://doi.org/10.1103/PhysRevE.82.046212} {\bibfield  {journal} {\bibinfo
  {journal} {Physical Review E}\ }\textbf {\bibinfo {volume} {82}},\ \bibinfo
  {pages} {046212} (\bibinfo {year} {2010})}\BibitemShut {NoStop}%
\bibitem [{\citenamefont {Soriano}\ \emph {et~al.}(2011)\citenamefont
  {Soriano}, \citenamefont {Zunino}, \citenamefont {Rosso}, \citenamefont
  {Fischer},\ and\ \citenamefont {Mirasso}}]{soriano2011}%
  \BibitemOpen
  \bibfield  {author} {\bibinfo {author} {\bibfnamefont {M.~C.}\ \bibnamefont
  {Soriano}}, \bibinfo {author} {\bibfnamefont {L.}~\bibnamefont {Zunino}},
  \bibinfo {author} {\bibfnamefont {O.~A.}\ \bibnamefont {Rosso}}, \bibinfo
  {author} {\bibfnamefont {I.}~\bibnamefont {Fischer}},\ and\ \bibinfo {author}
  {\bibfnamefont {C.~R.}\ \bibnamefont {Mirasso}},\ }\bibfield  {title}
  {\bibinfo {title} {Time scales of a chaotic semiconductor laser with optical
  feedback under the lens of a permutation information analysis},\ }\href
  {https://doi.org/10.1109/JQE.2010.2078799} {\bibfield  {journal} {\bibinfo
  {journal} {IEEE Journal of Quantum Electronics}\ }\textbf {\bibinfo {volume}
  {47}},\ \bibinfo {pages} {252} (\bibinfo {year} {2011})}\BibitemShut
  {NoStop}%
\bibitem [{\citenamefont {{De Micco}}\ \emph {et~al.}(2012)\citenamefont {{De
  Micco}}, \citenamefont {Fernández}, \citenamefont {Larrondo}, \citenamefont
  {Plastino},\ and\ \citenamefont {Rosso}}]{Micco2012}%
  \BibitemOpen
  \bibfield  {author} {\bibinfo {author} {\bibfnamefont {L.}~\bibnamefont {{De
  Micco}}}, \bibinfo {author} {\bibfnamefont {J.~G.}\ \bibnamefont
  {Fernández}}, \bibinfo {author} {\bibfnamefont {H.~A.}\ \bibnamefont
  {Larrondo}}, \bibinfo {author} {\bibfnamefont {A.}~\bibnamefont {Plastino}},\
  and\ \bibinfo {author} {\bibfnamefont {O.~A.}\ \bibnamefont {Rosso}},\
  }\bibfield  {title} {\bibinfo {title} {Sampling period, statistical
  complexity, and chaotic attractors},\ }\href
  {https://doi.org/10.1016/j.physa.2011.12.042} {\bibfield  {journal} {\bibinfo
   {journal} {Physica A: Statistical Mechanics and its Applications}\ }\textbf
  {\bibinfo {volume} {391}},\ \bibinfo {pages} {2564} (\bibinfo {year}
  {2012})}\BibitemShut {NoStop}%
\bibitem [{\citenamefont {Chrisment}\ and\ \citenamefont
  {Firpo}(2016)}]{Chrisment2016}%
  \BibitemOpen
  \bibfield  {author} {\bibinfo {author} {\bibfnamefont {A.~M.}\ \bibnamefont
  {Chrisment}}\ and\ \bibinfo {author} {\bibfnamefont {M.-C.}\ \bibnamefont
  {Firpo}},\ }\bibfield  {title} {\bibinfo {title} {Entropy–complexity
  analysis in some globally-coupled systems},\ }\href
  {https://doi.org/10.1016/j.physa.2016.05.009} {\bibfield  {journal} {\bibinfo
   {journal} {Physica A: Statistical Mechanics and its Applications}\ }\textbf
  {\bibinfo {volume} {460}},\ \bibinfo {pages} {162} (\bibinfo {year}
  {2016})}\BibitemShut {NoStop}%
\bibitem [{\citenamefont {Letellier}(2006)}]{Letellier2006}%
  \BibitemOpen
  \bibfield  {author} {\bibinfo {author} {\bibfnamefont {C.}~\bibnamefont
  {Letellier}},\ }\bibfield  {title} {\bibinfo {title} {Estimating the
  {S}hannon entropy: {R}ecurrence plots versus symbolic dynamics},\ }\href
  {https://doi.org/10.1103/PhysRevLett.96.254102} {\bibfield  {journal}
  {\bibinfo  {journal} {Physical Review Letters}\ }\textbf {\bibinfo {volume}
  {96}},\ \bibinfo {pages} {254102} (\bibinfo {year} {2006})}\BibitemShut
  {NoStop}%
\bibitem [{\citenamefont {Small}(2013)}]{Small2013}%
  \BibitemOpen
  \bibfield  {author} {\bibinfo {author} {\bibfnamefont {M.}~\bibnamefont
  {Small}},\ }\bibfield  {title} {\bibinfo {title} {Complex networks from time
  series: Capturing dynamics},\ }in\ \href
  {https://doi.org/10.1109/ISCAS.2013.6572389} {\emph {\bibinfo {booktitle}
  {2013 IEEE International Symposium on Circuits and Systems (ISCAS)}}}\
  (\bibinfo {year} {2013})\ pp.\ \bibinfo {pages} {2509--2512}\BibitemShut
  {NoStop}%
\bibitem [{\citenamefont {Masoller}\ \emph {et~al.}(2015)\citenamefont
  {Masoller}, \citenamefont {Hong}, \citenamefont {Ayad}, \citenamefont
  {Gustave}, \citenamefont {Barland}, \citenamefont {Pons}, \citenamefont
  {G{\'{o}}mez},\ and\ \citenamefont {Arenas}}]{Masoller2015}%
  \BibitemOpen
  \bibfield  {author} {\bibinfo {author} {\bibfnamefont {C.}~\bibnamefont
  {Masoller}}, \bibinfo {author} {\bibfnamefont {Y.}~\bibnamefont {Hong}},
  \bibinfo {author} {\bibfnamefont {S.}~\bibnamefont {Ayad}}, \bibinfo {author}
  {\bibfnamefont {F.}~\bibnamefont {Gustave}}, \bibinfo {author} {\bibfnamefont
  {S.}~\bibnamefont {Barland}}, \bibinfo {author} {\bibfnamefont {A.~J.}\
  \bibnamefont {Pons}}, \bibinfo {author} {\bibfnamefont {S.}~\bibnamefont
  {G{\'{o}}mez}},\ and\ \bibinfo {author} {\bibfnamefont {A.}~\bibnamefont
  {Arenas}},\ }\bibfield  {title} {\bibinfo {title} {{Quantifying sudden
  changes in dynamical systems using symbolic networks}},\ }\href
  {https://doi.org/10.1088/1367-2630/17/2/023068} {\bibfield  {journal}
  {\bibinfo  {journal} {New Journal of Physics}\ }\textbf {\bibinfo {volume}
  {17}},\ \bibinfo {pages} {023068} (\bibinfo {year} {2015})}\BibitemShut
  {NoStop}%
\bibitem [{\citenamefont {McCullough}\ \emph {et~al.}(2017)\citenamefont
  {McCullough}, \citenamefont {Small}, \citenamefont {Iu},\ and\ \citenamefont
  {Stemler}}]{McCullough2017}%
  \BibitemOpen
  \bibfield  {author} {\bibinfo {author} {\bibfnamefont {M.}~\bibnamefont
  {McCullough}}, \bibinfo {author} {\bibfnamefont {M.}~\bibnamefont {Small}},
  \bibinfo {author} {\bibfnamefont {H.~H.~C.}\ \bibnamefont {Iu}},\ and\
  \bibinfo {author} {\bibfnamefont {T.}~\bibnamefont {Stemler}},\ }\bibfield
  {title} {\bibinfo {title} {Multiscale ordinal network analysis of human
  cardiac dynamics},\ }\href {https://doi.org/10.1098/rsta.2016.0292}
  {\bibfield  {journal} {\bibinfo  {journal} {Philosophical Transactions of the
  Royal Society A: Mathematical, Physical and Engineering Sciences}\ }\textbf
  {\bibinfo {volume} {375}},\ \bibinfo {pages} {20160292} (\bibinfo {year}
  {2017})}\BibitemShut {NoStop}%
\bibitem [{\citenamefont {Leyva}\ \emph {et~al.}(2013)\citenamefont {Leyva},
  \citenamefont {Navas}, \citenamefont {Sendina-Nadal}, \citenamefont
  {Almendral}, \citenamefont {Buld{\'u}}, \citenamefont {Zanin}, \citenamefont
  {Papo},\ and\ \citenamefont {Boccaletti}}]{Leyva2013}%
  \BibitemOpen
  \bibfield  {author} {\bibinfo {author} {\bibfnamefont {I.}~\bibnamefont
  {Leyva}}, \bibinfo {author} {\bibfnamefont {A.}~\bibnamefont {Navas}},
  \bibinfo {author} {\bibfnamefont {I.}~\bibnamefont {Sendina-Nadal}}, \bibinfo
  {author} {\bibfnamefont {J.}~\bibnamefont {Almendral}}, \bibinfo {author}
  {\bibfnamefont {J.}~\bibnamefont {Buld{\'u}}}, \bibinfo {author}
  {\bibfnamefont {M.}~\bibnamefont {Zanin}}, \bibinfo {author} {\bibfnamefont
  {D.}~\bibnamefont {Papo}},\ and\ \bibinfo {author} {\bibfnamefont
  {S.}~\bibnamefont {Boccaletti}},\ }\bibfield  {title} {\bibinfo {title}
  {Explosive transitions to synchronization in networks of phase oscillators},\
  }\href {https://doi.org/10.1038/srep01281} {\bibfield  {journal} {\bibinfo
  {journal} {Scientific Reports}\ }\textbf {\bibinfo {volume} {3}},\ \bibinfo
  {pages} {1281} (\bibinfo {year} {2013})}\BibitemShut {NoStop}%
\bibitem [{\citenamefont {G{\'o}mez-Gardenes}\ \emph
  {et~al.}(2011)\citenamefont {G{\'o}mez-Gardenes}, \citenamefont {G{\'o}mez},
  \citenamefont {Arenas},\ and\ \citenamefont {Moreno}}]{Gardenes2011}%
  \BibitemOpen
  \bibfield  {author} {\bibinfo {author} {\bibfnamefont {J.}~\bibnamefont
  {G{\'o}mez-Gardenes}}, \bibinfo {author} {\bibfnamefont {S.}~\bibnamefont
  {G{\'o}mez}}, \bibinfo {author} {\bibfnamefont {A.}~\bibnamefont {Arenas}},\
  and\ \bibinfo {author} {\bibfnamefont {Y.}~\bibnamefont {Moreno}},\
  }\bibfield  {title} {\bibinfo {title} {Explosive synchronization transitions
  in scale-free networks},\ }\href
  {https://doi.org/10.1103/PhysRevLett.106.128701} {\bibfield  {journal}
  {\bibinfo  {journal} {Physical Review Letters}\ }\textbf {\bibinfo {volume}
  {106}},\ \bibinfo {pages} {128701} (\bibinfo {year} {2011})}\BibitemShut
  {NoStop}%
\bibitem [{\citenamefont {Leyva}\ \emph {et~al.}(2012)\citenamefont {Leyva},
  \citenamefont {Sevilla-Escoboza}, \citenamefont {Buld\'u}, \citenamefont
  {Sendi\~na Nadal}, \citenamefont {G\'omez-Garde\~nes}, \citenamefont
  {Arenas}, \citenamefont {Moreno}, \citenamefont {G\'omez}, \citenamefont
  {Jaimes-Re\'ategui},\ and\ \citenamefont {Boccaletti}}]{Leyva2012}%
  \BibitemOpen
  \bibfield  {author} {\bibinfo {author} {\bibfnamefont {I.}~\bibnamefont
  {Leyva}}, \bibinfo {author} {\bibfnamefont {R.}~\bibnamefont
  {Sevilla-Escoboza}}, \bibinfo {author} {\bibfnamefont {J.~M.}\ \bibnamefont
  {Buld\'u}}, \bibinfo {author} {\bibfnamefont {I.}~\bibnamefont {Sendi\~na
  Nadal}}, \bibinfo {author} {\bibfnamefont {J.}~\bibnamefont
  {G\'omez-Garde\~nes}}, \bibinfo {author} {\bibfnamefont {A.}~\bibnamefont
  {Arenas}}, \bibinfo {author} {\bibfnamefont {Y.}~\bibnamefont {Moreno}},
  \bibinfo {author} {\bibfnamefont {S.}~\bibnamefont {G\'omez}}, \bibinfo
  {author} {\bibfnamefont {R.}~\bibnamefont {Jaimes-Re\'ategui}},\ and\
  \bibinfo {author} {\bibfnamefont {S.}~\bibnamefont {Boccaletti}},\ }\bibfield
   {title} {\bibinfo {title} {Explosive first-order transition to synchrony in
  networked chaotic oscillators},\ }\href
  {https://doi.org/10.1103/PhysRevLett.108.168702} {\bibfield  {journal}
  {\bibinfo  {journal} {Physical Review Letters}\ }\textbf {\bibinfo {volume}
  {108}},\ \bibinfo {pages} {168702} (\bibinfo {year} {2012})}\BibitemShut
  {NoStop}%
\bibitem [{\citenamefont {Letellier}\ and\ \citenamefont
  {Aguirre}(2002)}]{Letellier2002}%
  \BibitemOpen
  \bibfield  {author} {\bibinfo {author} {\bibfnamefont {C.}~\bibnamefont
  {Letellier}}\ and\ \bibinfo {author} {\bibfnamefont {L.~A.}\ \bibnamefont
  {Aguirre}},\ }\bibfield  {title} {\bibinfo {title} {Investigating nonlinear
  dynamics from time series: The influence of symmetries and the choice of
  observables},\ }\href {https://doi.org/10.1063/1.1487570} {\bibfield
  {journal} {\bibinfo  {journal} {Chaos}\ }\textbf {\bibinfo {volume} {12}},\
  \bibinfo {pages} {549} (\bibinfo {year} {2002})}\BibitemShut {NoStop}%
\bibitem [{\citenamefont {Sendi{\~n}a-Nadal}\ and\ \citenamefont
  {Letellier}(2022)}]{Sendina2022}%
  \BibitemOpen
  \bibfield  {author} {\bibinfo {author} {\bibfnamefont {I.}~\bibnamefont
  {Sendi{\~n}a-Nadal}}\ and\ \bibinfo {author} {\bibfnamefont {C.}~\bibnamefont
  {Letellier}},\ }\bibfield  {title} {\bibinfo {title} {Observability analysis
  and state reconstruction for networks of nonlinear systems},\ }\href
  {https://doi.org/10.1063/5.0090239} {\bibfield  {journal} {\bibinfo
  {journal} {Chaos}\ }\textbf {\bibinfo {volume} {32}},\ \bibinfo {pages}
  {083109} (\bibinfo {year} {2022})}\BibitemShut {NoStop}%
\bibitem [{\citenamefont {Chialvo}(1995)}]{chialvo1995}%
  \BibitemOpen
  \bibfield  {author} {\bibinfo {author} {\bibfnamefont {D.~R.}\ \bibnamefont
  {Chialvo}},\ }\bibfield  {title} {\bibinfo {title} {Generic excitable
  dynamics on a two-dimensional map},\ }\href
  {https://doi.org/10.1016/0960-0779(93)E0056-H} {\bibfield  {journal}
  {\bibinfo  {journal} {Chaos, Solitons \& Fractals}\ }\textbf {\bibinfo
  {volume} {5}},\ \bibinfo {pages} {461} (\bibinfo {year} {1995})}\BibitemShut
  {NoStop}%
\bibitem [{\citenamefont {Stankevich}\ \emph {et~al.}(2023)\citenamefont
  {Stankevich}, \citenamefont {Gonchenko}, \citenamefont {Popova},\ and\
  \citenamefont {Gonchenko}}]{Stankevich2023}%
  \BibitemOpen
  \bibfield  {author} {\bibinfo {author} {\bibfnamefont {N.~V.}\ \bibnamefont
  {Stankevich}}, \bibinfo {author} {\bibfnamefont {A.~S.}\ \bibnamefont
  {Gonchenko}}, \bibinfo {author} {\bibfnamefont {E.~S.}\ \bibnamefont
  {Popova}},\ and\ \bibinfo {author} {\bibfnamefont {S.~V.}\ \bibnamefont
  {Gonchenko}},\ }\bibfield  {title} {\bibinfo {title} {Complex dynamics of the
  simplest neuron model: Singular chaotic shilnikov attractor as specific
  oscillatory neuron activity},\ }\href
  {https://doi.org/10.1016/j.chaos.2023.113565} {\bibfield  {journal} {\bibinfo
   {journal} {Chaos, Solitons \& Fractals}\ }\textbf {\bibinfo {volume}
  {172}},\ \bibinfo {pages} {113565} (\bibinfo {year} {2023})}\BibitemShut
  {NoStop}%
\bibitem [{\citenamefont {Boccaletti}\ \emph {et~al.}(2002)\citenamefont
  {Boccaletti}, \citenamefont {Kurths}, \citenamefont {Osipov}, \citenamefont
  {Valladares},\ and\ \citenamefont {Zhou}}]{Boccaletti2002}%
  \BibitemOpen
  \bibfield  {author} {\bibinfo {author} {\bibfnamefont {S.}~\bibnamefont
  {Boccaletti}}, \bibinfo {author} {\bibfnamefont {J.}~\bibnamefont {Kurths}},
  \bibinfo {author} {\bibfnamefont {G.}~\bibnamefont {Osipov}}, \bibinfo
  {author} {\bibfnamefont {D.}~\bibnamefont {Valladares}},\ and\ \bibinfo
  {author} {\bibfnamefont {C.}~\bibnamefont {Zhou}},\ }\bibfield  {title}
  {\bibinfo {title} {The synchronization of chaotic systems},\ }\href
  {https://doi.org/10.1016/S0370-1573(02)00137-0} {\bibfield  {journal}
  {\bibinfo  {journal} {Physics Reports}\ }\textbf {\bibinfo {volume} {366}},\
  \bibinfo {pages} {1} (\bibinfo {year} {2002})}\BibitemShut {NoStop}%
\bibitem [{\citenamefont {R{\"o}ssler}(1976)}]{Ros76}%
  \BibitemOpen
  \bibfield  {author} {\bibinfo {author} {\bibfnamefont {O.~E.}\ \bibnamefont
  {R{\"o}ssler}},\ }\bibfield  {title} {\bibinfo {title} {An equation for
  continuous chaos},\ }\href {https://doi.org/10.1016/0375-9601(76)90101-8}
  {\bibfield  {journal} {\bibinfo  {journal} {Physics Letters A}\ }\textbf
  {\bibinfo {volume} {57}},\ \bibinfo {pages} {397} (\bibinfo {year}
  {1976})}\BibitemShut {NoStop}%
\end{thebibliography}
\end{document}